\documentclass[structabstract]{aa}  
\usepackage{graphicx}
\usepackage{txfonts}
\usepackage{natbib}
\bibpunct{(}{)}{;}{a}{}{,}

\begin{document}
   \title{The AMIGA sample of isolated galaxies}

   \subtitle{XI. Optical characterisation of nuclear activity\fnmsep\thanks{Full
tables~\ref{table:add}-\ref{table:class}, \ref{table:phot}-\ref{table:hl_hcg}
are only available in electronic form at http://amiga.iaa.es/ and at the CDS via
anonymous ftp to cdsarc.u-strasbg.fr (130.79.128.5)
or via http://cdsweb.u-strasbg.fr/cgi-bin/qcat?J/A+A/545/A15}}

   \author{J. Sabater
          \inst{1,2}
          \and
          L. Verdes-Montenegro\inst{1}
          \and
          S. Leon\inst{3}
          \and
          P. Best\inst{2}
          \and
          J. Sulentic\inst{1}      
          }

   \institute{Instituto de Astrof\'{\i}sica de Andaluc\'{\i}a, CSIC,
              Apdo. 3004, 18080
              Granada, Spain\\
              \email{jsm@iaa.es}
         \and
             Institute for Astronomy, University of Edinburgh, 
             Royal Observatory, Blackford Hill,
             EH9 3HJ Edinburgh, UK
         \and
              ALMA, European Southern Observatory, 
              Alonso de C\'ordova, 3107, Santiago,
              Chile\\
              }

   \offprints{J. Sabater}

   \date{Received 20 December, 2011; accepted 29 May, 2012}

% \abstract{}{}{}{}{} 
% 5 {} token are mandatory
 
  \abstract
   {This paper is part of a series involving the AMIGA project (Analysis of the 
Interstellar Medium of Isolated GAlaxies), which identifies and studies a 
statistically significant sample of the most isolated galaxies in the northern 
sky.}
   {We present a catalogue of nuclear activity, traced by optical
emission lines, in a well-defined sample of the
most isolated galaxies in the local Universe, 
which will be used as a basis for studying the effect of the environment
on nuclear activity.}
   {We obtained spectral data from the 6th Data Release of the
Sloan Digital Sky Survey, which were inspected in a semi-automatic way. We
subtracted the underlying stellar populations from the spectra (using
the software Starlight) and modelled the nuclear emission features. Standard
emission-line
diagnostics
diagrams were applied, using a new classification scheme that takes into account
censored data, to classify the type of nuclear emission.}
   {We provide a final catalogue of spectroscopic data, stellar populations,
emission lines and classification of optical nuclear activity for AMIGA
galaxies. The prevalence of optical active galactic nuclei (AGN) in AMIGA
galaxies is 20.4\%, or 36.7\% including
transition objects. The fraction of AGN increases steeply towards earlier
morphological types and higher luminosities. We compare these results
with a
matched analysis of galaxies in isolated denser environments 
(Hickson Compact Groups). After correcting for the
effects of the morphology and luminosity, we find that there is no evidence for
a difference in
the prevalence of AGN between isolated and compact group galaxies, and we
discuss the implications of this result.}
   {We find that a major interaction is not a necessary condition for the
triggering of optical AGN.
}


   \maketitle
%
%________________________________________________________________

\section{Introduction}

It is widely accepted that galaxy evolution is strongly influenced, or even
driven, by environment \citep[nurture; see e.g. ][and references
therein]{Park2009,Liu2012}. Galaxy interactions are thought to induce nuclear
activity by removing angular momentum from the gas and, in this way, feeding the
central black hole \citep{Shlosman1990,Barnes1991,Haan2009,Liu2011}. Hence, a
higher rate of active galactic nuclei (AGN) is expected among interacting
galaxies. 

However, different studies have yielded contradictory results: some studies find
a higher local density of companions near galaxies hosting an AGN or a higher
prevalence of AGN in interacting galaxies
\citep{Petrosian1982,Dahari1985,MacKenty1990,Rafanelli1995,
Rafanelli1997,Alonso2007,Ellison2011,Liu2012}, while others find no excess, or
only a marginal excess
\citep{Bushouse1986,Laurikainen1995,Schmitt2001a,Miller2003,Ellison2008,Li2008}.
\citet{Schawinski2009,Schawinski2010} found a clear relation between some AGN
and the signatures of a previous merger. In recent studies the relation between
X-Ray AGN and environment was also explored
\citep{Silverman2009,Haggard2010,Tasse2011}, finding that this relation may
depend on the mass of the host galaxy \citep{Silverman2009}. Another possible
source of discrepancy involves the methodology used to establish the presence of
an AGN. Subtraction or non-subtraction of the stellar component from the nuclear
light may lead to differences in the nuclear emission classification. This is
particularly true for weak emission lines, which might be affected not just by a
low signal-to-noise ratio, but also by details of the stellar spectrum
\citep{Ho1997}. The use of different classification criteria to evaluate 
whether the nuclear emission is powered by star formation (SF) or by an AGN also
complicates the direct comparison of results found in different studies.

The AMIGA project \citep[Analysis of the interstellar Medium in Isolated
GAlaxies; http://amiga.iaa.es/]{Verdes-Montenegro2005} was initiated to separate
the effect of galaxy interactions from the intrinsic evolution in a galaxy.
AMIGA provides a panchromatic characterisation of a well-defined and
statistically significant sample of isolated galaxies. AMIGA can be described as
a vetted or value-added catalogue, based upon the Catalogue of Isolated Galaxies
\citep[CIG;][]{Karachentseva1973}, which is composed of 1050 galaxies compiled
using an isolation criterion that implies that the galaxies have probably been
unperturbed for $\approx$ 3 Ga \citep{Verdes-Montenegro2005}\footnote{We will
follow the recommendations for units of the IAU Style Manual
\citep{Wilkins1995}. Hence, we use the term annus, abbreviated as ``a'', for
year (a - annus - year; Ma - Megaannus - Megayear; Ga - Gigaannus - Gigayear).}.
Dwarf companions (B$>$-18 to -15 depending on distance) are not (cannot be)
excluded, but the degree of isolation from major companions (greater than 10\%
of the mass of the primary) was re-evaluated and quantified for each galaxy in
terms of both the local number density of neighbours and tidal strength
\citep{Verley2007a, Verley2007b}.

Multiwavelength studies of the AMIGA sample (data released via a Virtual
Observatory interface) have placed special emphasis on structural parameters as
well as different components/phases of the interstellar medium. Our studies
suggest that the most isolated galaxies show different properties than less
isolated field samples. AMIGA early-type galaxies are usually fainter than
late-types in the B-band and most spirals in our sample appear to host
pseudo-bulges and not classical bulges
\citep{Verdes-Montenegro2005,Sulentic2006,Durbala2008}. AMIGA spiral galaxies
are redder than similar type galaxies in close pairs, showing a Gaussian
distribution of the (g$-$r) colours with a smaller median absolute deviation
(almost half) compared to galaxies in wide and close pairs. Properties such as
far-infrared (FIR) and radio continuum emission, which usually show enhancement
in interacting galaxy samples \citep{Darg2010}, show levels in AMIGA that are at
or below those measured in any other galaxy samples: This has been shown for,
$L_{\mathrm{FIR}}$ \citep{Lisenfeld2007}, $L_{1.4\mathrm{GHz}}$
\citep{Leon2008}, radio-excess above the radio-FIR correlation
\citep[0\%;][]{Sabater2008,TesisMia}, the \ion{H}{i} asymmetry
\citep{Espada2011b}, and the molecular gas content \citep{Lisenfeld2011}. AMIGA
galaxies are also located in the lowest environmental densities
\citep{Verley2007b} and do not present any morphological signatures of
interactions \citep{Sulentic2006}.

Here, we present a catalogue of nuclear properties for AMIGA galaxies obtained
from optical spectra provided by the Sloan Digital Sky Survey
\citep[SDSS;][]{York2000,Adelman-McCarthy2008} in its 6th Data Release. In
Sect.~\ref{sec:data} we present the sample of galaxies used in this study and 
describe the compilation and reduction of the data, including the subtraction of
the stellar contribution from the nuclear spectra and the measurement of
emission lines. The method used for the classification of the nuclei, which
takes into account the presence of censored data,  is presented in
Sect.~\ref{sec:class}, together with the final classification of the nuclei and
its relation with the properties of the galaxies. A comparison with galaxies in
a denser environment (compact groups) is analysed in Sect.~\ref{sec:comp}.
Finally, we present a summary of the work, followed by a discussion about the
effect of major-mergers in the triggering of nuclear activity and our
conclusions, in Sect.~\ref{sec:discon}

%__________________________________________________________________

\section{Data}
\label{sec:data}

\subsection{The sample}
\label{sec:sample}

We adopted the AMIGA sample of isolated galaxies as the nurture-free base
sample for our study. The isolation criteria for AMIGA galaxies imply
nearest-neighbour crossing times of $\approx 3$ Ga or more \citep[][in
prep.]{Verdes-Montenegro2005,Verley2007b,Mamen2012}. This, and the properties of
the sample presented in the introduction, lead us to hypothesize that AMIGA
galaxies were not affected by major tidal interactions during the last part of
their lifetime.

The SDSS provides a large homogeneous database of optical spectra from which
line ratios can be extracted and quantified to investigate nuclear activity. We
used Data Release 6 (DR6), which was the one available when we started the
study. We found a total of 549 AMIGA galaxies in the SDSS-DR6 database
\citep{Adelman-McCarthy2008} with nuclear spectra for 362 of the galaxies. We
found that spectra for 353 galaxies could be used for the present study as
explained in Sect.~\ref{sec:sdss_spec}. Henceforth, we refer to this subset of
$n = 353$ isolated galaxies as \textit{the SDSS sample}.

\begin{figure}
\centering
 \resizebox{8.5cm}{8.5cm}{\includegraphics{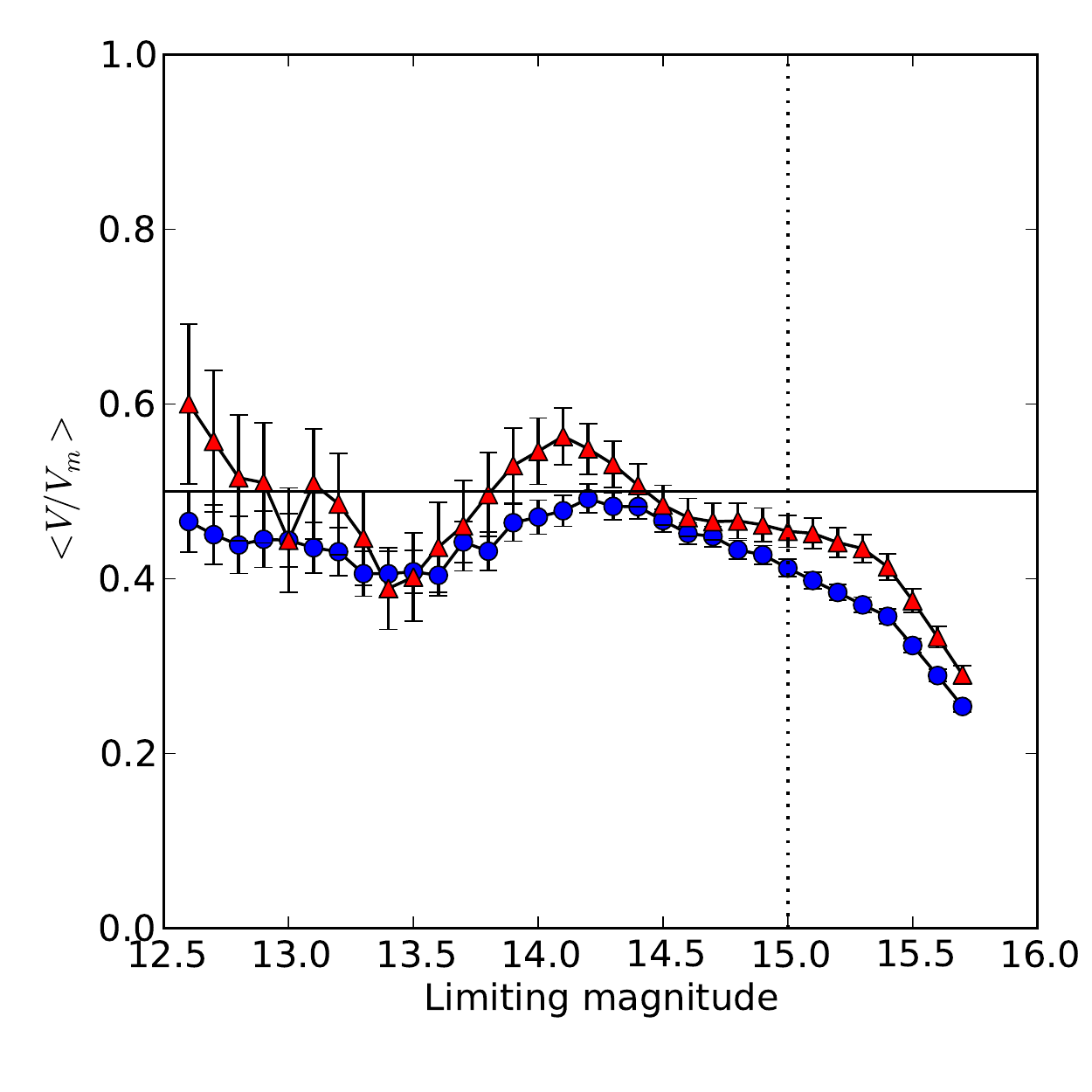}}
   \caption{Result of a $\left\langle V/V_{m}\right\rangle$ completeness test 
for the CIG SDSS sample ($n=353$, red triangles) and the CIG full sample
($n=1050$, blue circles). 
           }
      \label{fig:sdss_compl}
\end{figure}

We estimated the completeness of the SDSS sample using a $\left\langle
V/V_{m}\right\rangle$ test \citep[as in][]{Verdes-Montenegro2005} and found it
to be even more complete than the full AMIGA sample up to a limiting magnitude
of $m_B = 15.0$ (see Fig.~\ref{fig:sdss_compl}). This reflects the completeness
and uniformity of data within the region of sky covered by SDSS. For studies
that require statistical significance and completeness we selected those
galaxies with $m_B$ in the range 11.0 to 15.0 \citep{Lisenfeld2007}. We excluded
galaxies flagged as interacting in \citet[marked with the parameter I/A =
yes]{Sulentic2006} as well as local dwarf galaxies CIG~663 (UMi dwarf) and
CIG~802 (Draco dwarf). We refer to this sample, composed of $n = 226$ galaxies,
as \textit{the SDSS complete sample}.

\subsection{Morphology and luminosity}
\label{sec:lt}

To study the relation of the nuclear activity type with the properties of the
host galaxies we also used morphological data, optical luminosity and
near-infrared (NIR) luminosity. The morphological data are based on work
presented in \citet{Sulentic2006} and \citet{Mirian2012}. Optical luminosities
($L_{\mathrm{B}}$) were presented in \citet{Verdes-Montenegro2005} and were
refined based on new distances for AMIGA galaxies and a careful revision of the
velocities and corrections \citep{Mirian2012}. The infrared luminosity was
derived from the \mbox{$\mathrm{K_s}$-band} (2.159 $\mathrm{\mu m}$) magnitude
obtained from the Two Micron All Sky Survey (2MASS) extended source catalogue
\citep{Jarrett2000}. The $L_{\rm{K}}$ was obtained from the total (extrapolated)
$\mathrm{K_s}$ flux considering that $L_{\rm{K}} = 4\pi d^2 \nu
f_{\mathrm{K_s}}(\nu)$, where $d$ the distance to the galaxy, $\nu$ the central
frequency of the K$_{\mathrm{s}}$ band and $f_{\mathrm{K_s}}(\nu)$ the flux
density derived from the magnitude in the catalogue. Finally, the luminosity was
normalised by the solar luminosity in K$_{\mathrm{s}}$ band
(L$_{\mathrm{K,\sun}}$). After a visual inspection we found some galaxies
associated with two 2MASS sources within our search radius. In each case, the
flux of the brighter one is the one corresponding to the whole galaxy, hence it
was selected for these galaxies \citep{Lisenfeld2011}. Although a
mass-to-luminosity ratio could be applied to derive a total stellar mass
\citep{Cole2001}, it depends on the star formation history of the galaxy and on
the initial mass function considered \citep{Bell2001,Bell2003}. Hence, we
directly used $L_{\mathrm{K}}$ as an estimator of the stellar mass content of
the galaxy. The optical and infrared luminosities, as well as the morphological
classification, are presented in Table~\ref{table:add}.

\begin{table*}
\caption{Near-infrared and optical luminosities and morphologies of CIG
galaxies.$^{1}$}
\label{table:add}
~\\
\centering
\begin{tabular}{c c c c c c}
\hline
CIG & $f_{\mathrm{K_s}}$ $^{2}$& $\log(L_{\mathrm{K}})$ $^{2}$ & code$^3$ &
$\log(L_{\mathrm{B}})$ $^{2,4}$ & morpho.$^4$ \\
      &  Jy &  [L$_{\mathrm{K,\sun}}$]   &         & [L$_{\sun}$]  &  RC3 \\
\hline \hline
1& $0.0548 \pm 0.0020$ & $11.227 \pm 0.016$ &1&10.57& $5.0 \pm 1.5$ \\
2& $0.0087 \pm 0.0010$ & $10.407 \pm 0.050$ &1&9.84& $6.0 \pm 1.5$ \\
3& $0.00539 \pm 0.00053$ &$-$&$-1$&$-$& $4.0 \pm 1.5$ \\
4& $0.2855 \pm 0.0021$ & $10.9777 \pm 0.0032$ &1&10.28& $3.0 \pm 1.5$\\
5& $0.01320 \pm 0.00043$ & $10.686 \pm 0.014$ &1&9.95& $0.0 \pm 1.5$\\
6& $0.0129 \pm 0.0013$ & $10.205 \pm 0.042$ &1&9.80& $7.0 \pm 1.5$ \\
7& $0.0197 \pm 0.0012$ & $11.270 \pm 0.027$ &1&10.35& $4.0 \pm 1.5$ \\
 ... & ... & ... & ... & ... & ... \\
\hline
\end{tabular}
\begin{list}{}{}
 \item $^1$ AMIGA sample ($n = 1050$).
 \item $^2$ Units: 
$\log(L_{\mathrm{K}})$ given in units of \mbox{$\mathrm{K_s}$-band} solar
luminosity (L$_{\mathrm{K,\sun}} = 5.0735 \times 10^{25}
\mathrm{W}$); 
$\log(L_{\mathrm{B}})$ given in units of solar bolometric
luminosity (L$_{\sun} = 3.842 \times 10^{26}
\mathrm{W}$); $1
\mathrm{Jy} = 10^{-26} \mathrm{W} \mathrm{m^{-2}} \mathrm{Hz^{-1}}$.
 \item $^3$ Code for the infrared luminosity: 
 $-1$ if the galaxy does not have distance data;
 0 if the galaxy does not have NIR data;
 1 if the galaxy has NIR data and distance;
 2 if there are 2 NIR sources for the galaxy, the brightest one is selected (see
text).
 \item $^4$ $\log(L_{\mathrm{B}})$ and morphology data from
\citet{Mirian2012}.
\end{list}
\end{table*}

Additional data on the nuclear activity of AMIGA galaxies were presented in
\citet{Sabater2008} and \citet{TesisMia}. The classifications in those papers
were taken from the literature, and obtained from the study of the radio-FIR
correlation and the FIR colour.

\begin{figure}
\centering
\includegraphics[width=0.5\textwidth]{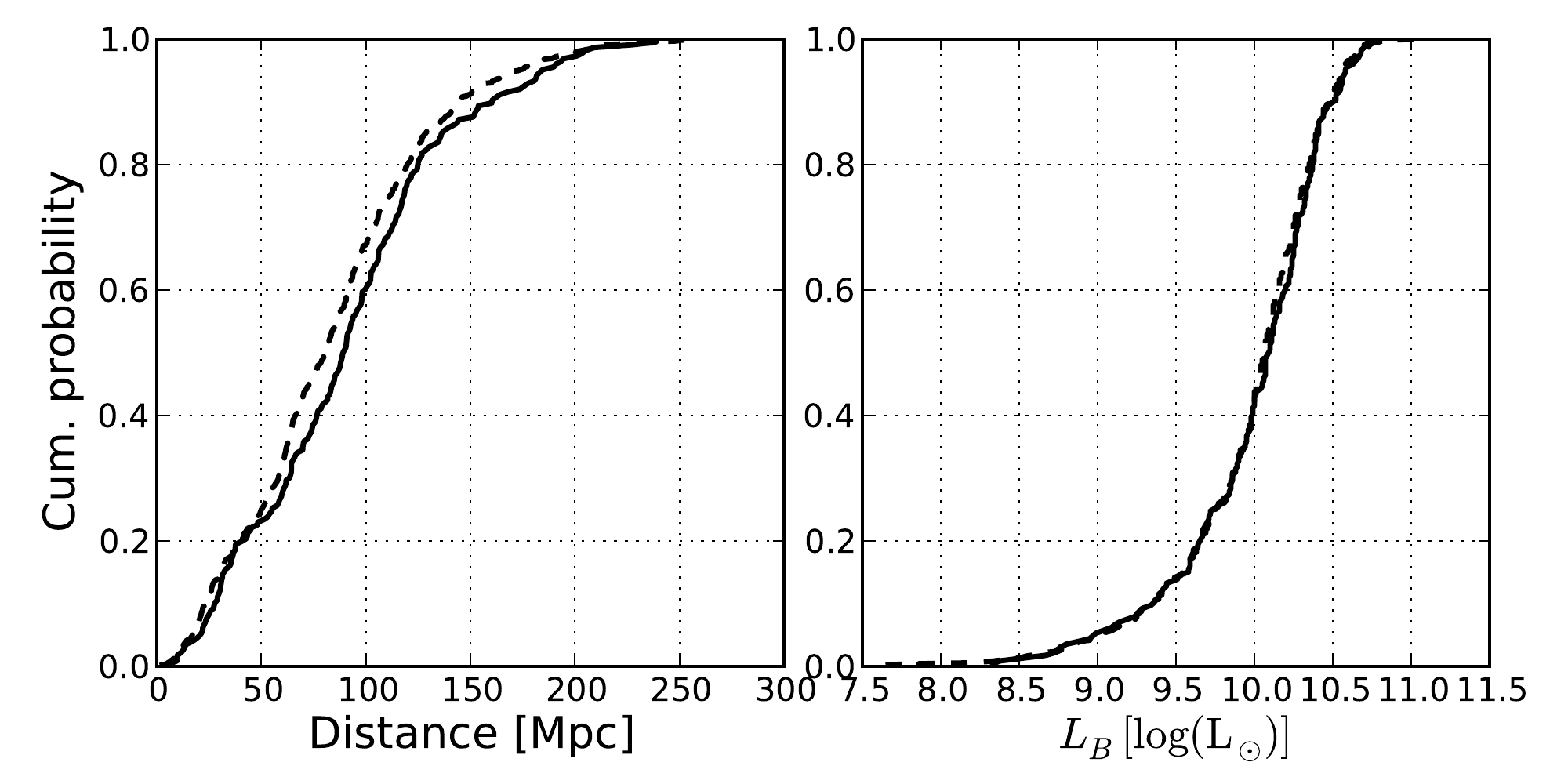}
   \caption{Cumulative distributions of distance (left) and optical 
luminosity (right) for the AMIGA (solid line) and SDSS complete samples 
(dashed line) used in the Kolmogorov-Smirnov two-sample test.
           }
      \label{fig:comb_stats}
\end{figure}

\begin{figure}
\centering
\includegraphics[width=0.4\textwidth]{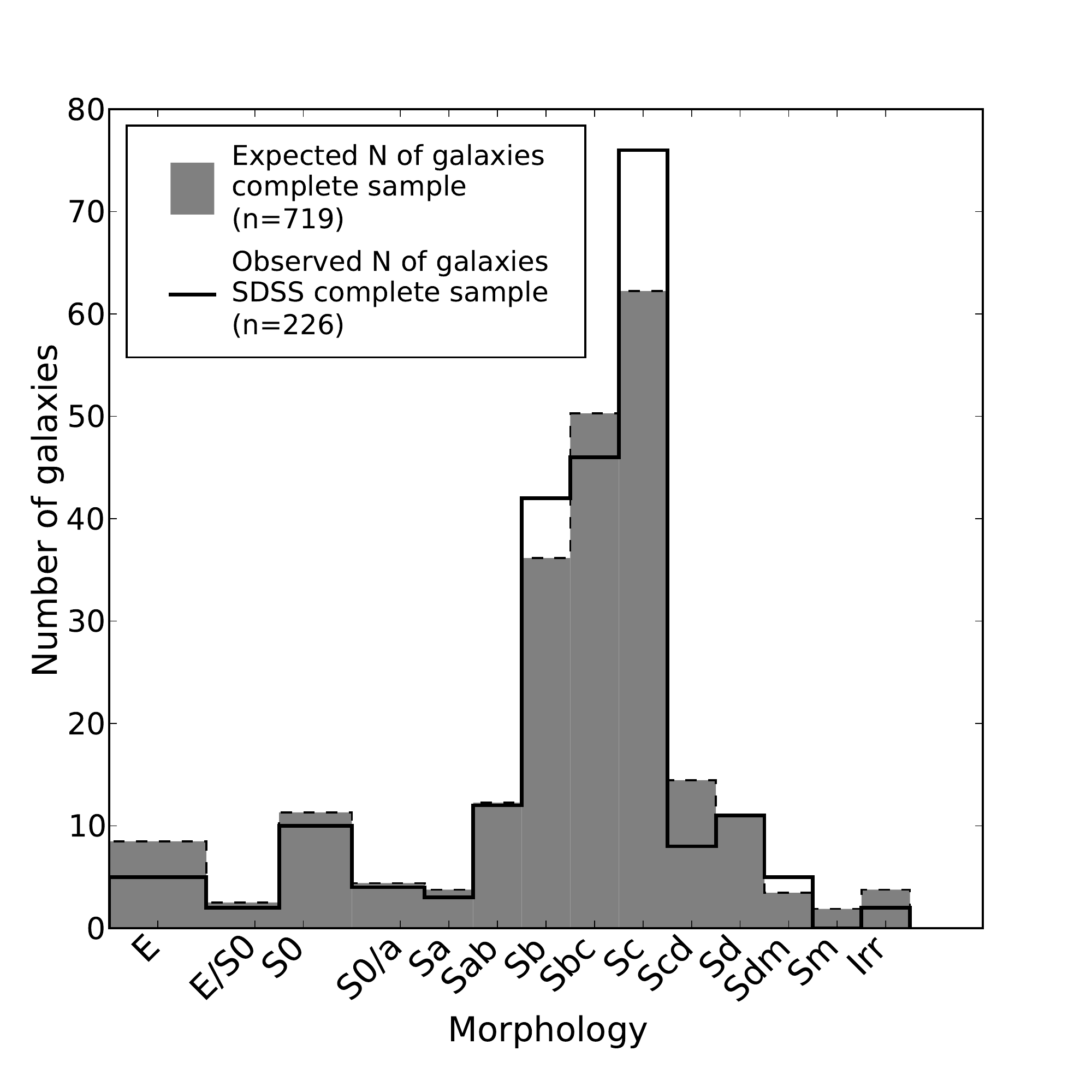}
\caption{Binned distributions of morphological types for the observed (open
bars) frequencies for the SDSS complete sample and scaled (grey-filled bars)
frequencies derived from the CIG complete sample. These frequencies were 
used to perform a $\chi^{2}$ test.}
\label{fig:chi2_morfo}
\end{figure}

We tested whether the SDSS complete sample represents well some fundamental
properties of its parent sample \citep[the AMIGA complete sample,
][]{Lisenfeld2007}. In Fig.~\ref{fig:comb_stats} the cumulative distributions of
distances and luminosities are shown (left and right panels, respectively) for
the AMIGA complete sample and the SDSS complete sample. To quantify the
comparisons we used a Kolmogorov-Smirnov two-sample test
\citep[K-S;][]{Kolmogorov1933,Smirnov1936} to estimate the likelihood that the
SDSS and AMIGA samples are derived from the same (undefined) parent population.
The K-S test significance levels are 0.15 and 0.69 for comparisons of distance
and source luminosity, respectively. These values are sufficiently high to
support the hypothesis that the two samples come from the same parent population
at a 10\% significance level ($\alpha = 0.1$, which is widely used in the
literature). We also applied a $\chi^{2}$ test to compare the morphology
distribution of the two samples. This test is especially suitable in this case
because of the discrete nature of the morphological classification.
Fig.~\ref{fig:chi2_morfo} presents the morphological distribution of
classifications. As we can see in the figure, and as the $\chi^{2}$ test
confirms, the distributions are very similar in all cases. The significance
level is higher than $\alpha = 0.10$ (p-value = 0.484). Therefore, we have no
reason to reject the hypothesis that the SDSS sample follows the same
distribution as the parent sample at a significance level of 10\%. We therefore
conclude that no clear bias exists in the SDSS complete sample relative to the
AMIGA complete sample and  we consider the SDSS complete sample used in this
study to be representative of the AMIGA complete sample.

\subsection{SDSS spectroscopic data}
\label{sec:sdss_spec}

Spectra were found for $n=398$ AMIGA galaxies in the SDSS DR6 catalogue.
Nuclear spectra were available for 362 of these galaxies and non-nuclear
spectra were discarded for our study. We downloaded the corresponding
spectra and compiled the flux density, its error and the error mask for each
wavelength. 

We extracted measures using an alternative reduction pipeline of the SDSS data
(called specBS). This Princeton University product is publicly available at
\textit{http://spectro.princeton.edu/} where the spectral parameters are
determined via a comparison with carefully selected templates of stars, galaxies
and quasars from the SDSS and ELODIE surveys \citep{Moultaka2004}. This resulted
in improved measures of the velocity dispersion and the redshift. The velocity
dispersion could be reliably determined for 42\% of the galaxies in our sample
using the standard pipeline and 98\% of the galaxies using the Princeton
pipeline. Redshift discrepancies  between the two procedures were negligible
(there are only three galaxies with a difference of more than 100 km/s) although
the uncertainty in the redshift measurements is lower using the Princeton
pipeline. As a final test, we visually checked all spectra and found that some
of them had regions in which spectral data was missing. In some of the cases the
spectral gaps did not affect the regions of interest for this study. In only one
case (CIG~347) the entire spectral data were corrupted.

\label{ref:comments}
After flagging galaxies with bad spectral regions, we have $n = 353$ galaxies
for our study of nuclear spectra in the AMIGA sample. The spectroscopic data are
presented in Table~\ref{table:spec}. In the column ``Comments'' we marked
galaxies with bad spectral data. These nine flagged galaxies were not included
in our study to avoid inhomogeneities in the data sample. K-S two sample tests
were performed among the raw sample, the clean sample and the flagged galaxies
for the distance, optical luminosity and B magnitude. Throughout, the p-values
were higher than 0.8. A $\chi^{2}$ was applied for the morphology and a p-value
of 0.99 was found. The properties (luminosity, morphology and distance) of the
flagged galaxies are comparable to those of the remaining galaxies;
consequently, no bias is expected to be introduced by removing them.

\begin{table*}
\caption{Catalogue of SDSS nuclear spectra and some of their properties.$^1$}
\label{table:spec}
~\\
\centering
\begin{tabular}{r r r r c c c c c c}
\hline
CIG & plate & mjd~~ & fiberId & Compl.$^2$ & z & V. disp. & z (specBS) &
V. disp. (specBS) & Comm.$^3$ \\
    &       &     &         &        &   & km/s &            & km/s &  \\
\hline \hline
11 & 389 & 51795 & 208 & 1 & $0.01316 \pm 0.00012$ & - & $0.013181 \pm 0.000013$ & $49.6 \pm 6.0$ &  \\
16 & 390 & 51900 & 522 & 0 & $0.01824 \pm 0.00015$ & $103.0 \pm 2.0$ & $0.0182119 \pm 0.0000054$ & $102.2 \pm 2.0$ &  \\
19 & 753 & 52233 & 95 & 0 & $0.01751 \pm 0.00018$ & $169.0 \pm 5.0$ & $0.0175050 \pm 0.0000090$ & $174.6 \pm 3.8$ &  \\
56 & 400 & 51820 & 582 & 1 & $0.017262 \pm 0.000099$ & - & $0.0172053 \pm 0.0000044$ & $134.6 \pm 4.5$ &  \\
60 & 401 & 51788 & 223 & 1 & $0.01727 \pm 0.00011$ & - & $0.017191 \pm 0.000014$ & $31 \pm 19$ &  \\
187 & 1864 & 53313 & 171 & 1 & $0.02767 \pm 0.00016$ & $77.0 \pm 5.0$ & $0.027633 \pm 0.000011$ & $86.0 \pm 3.9$ &  \\
189 & 1865 & 53312 & 417 & 1 & $0.01046 \pm 0.00019$ & $191.0 \pm 4.0$ & $0.0104411 \pm 0.0000067$ & $189.9 \pm 3.0$ & \\
 ... & ... & ... & ... & ... & ... & ... & ... & ... & ...\\
\hline
\end{tabular}
\begin{list}{}{}
 \item $^1$ AMIGA galaxies with SDSS DR6 nuclear spectra ($n=362$).
Columns: 
(1) CIG catalogue number; 
(2) SDSS plate; 
(3) SDSS modified Julian date; 
(4) SDSS fiber Id; 
(5) code for membership in the complete sample;$^2$ 
(6) redshift and its uncertainty; 
(7) velocity dispersion and uncertainty; 
(8) redshift from the specBS pipeline and uncertainty; 
(9) velocity dispersion from the specBS pipeline and uncertainty and 
(10) comments.
 \item $^2$ Code of completeness: 1 if the galaxy belongs to the SDSS complete 
sample and 0 otherwise.
 \item $^3$ Comments: ``BL'' for galaxies with broad lines (see 
Sec.~\ref{sec:broadlines}) and ``BZ'' for galaxies with bad spectral zones (see
text, Sect.~\ref{ref:comments}).
\end{list}
\end{table*}

\subsection{Stellar populations}

The spectrum of a galaxy includes contributions from both stellar (absorption)
and gaseous (emission) line components. Both are of potential interest but they
must be separated to draw any quantitative conclusions. The stellar absorption
spectrum must be extracted to perform a reliable fitting of the emission lines
\citep{Ho1997}. Most methods used to determine the stellar populations are based
on templates of stellar absorption-line spectra that are used as the basis to
fit  the stellar component
\citep[e.g.][]{Ho1997,Engelbracht1998,Kauffmann2003a,CidFernandes2005,Hao2005}.
The library of templates should be large to allow the fitting of stellar
templates that reflect a wide dispersion in galaxy properties such as
metallicities, ages and velocity dispersions. From among the available codes we
selected Starlight \citep{CidFernandes2005}, not only because of its ability to
accurately determine the properties of the underlying populations, but also
because it is publicly available, easy to use, and well documented.

Starlight is designed to fit an observed spectrum with a model
\citep{CidFernandes2005,Mateus2006,Asari2007} composed of spectral components
from a pre-defined set of input spectra. The spectral base can be made up of
observed template spectra, evolutionary synthesis models
\citep{CidFernandes2004,CidFernandes2005}, individual stars, etc. We used 45
synthetic spectra from \citet{Bruzual2003} as the spectral base, with three
metallicities (Z) and 15 different ages. The 3 metallicities are: 0.004 (0.2
Z$_{\sun}$), 0.02 (Z$_{\sun}$) and 0.05 (2.5 Z$_{\sun}$) and the 15 ages are:
1.0 Ma, 3.16 Ma, 5.01 Ma, 10.0 Ma, 25.12 Ma, 40.0 Ma, 101.52 Ma, 286.12 Ma,
640.54 Ma, 904.79 Ma, 1.434 Ga, 2.5 Ga, 5.0 Ga, 11.0 Ga and 13 Ga. These base
spectra are distributed together with Starlight and reliably fit the stellar
populations of typical SDSS galaxies \citep{CidFernandes2005}. The age-Z
degeneracy is estimated to slightly bias the estimation of the average of Z and
log(age) at the level of up to $\sim 0.1-0.2$ dex
\citep{CidFernandes2005,Cid-Fernandes2010b,GonzalezDelgado2010}.

We developed a software to adapt the original SDSS spectra to the input of
Starlight, which performed the following tasks: a) uniform re-sampling of the
spectra with a final resolution of 0.1~nm and transformation from vacuum to air
wavelengths based on IAU standards \citep{Morton1991}; b) dereddening of the
flux and its uncertainty based on the dust emission maps of \citet{Schlegel1998}
and the extinction curve by \citet{O'Donnell1994}, which is an improvement of
the extinction curve of \citet{Cardelli1989} in the optical and the near
infrared; and c) shifting of the spectra to the rest-frame using the redshifts
from the Princeton SDSS data (specBS). The initial values of the velocity
dispersions were also extracted from the Princeton data (see
Table~\ref{table:spec}).

\begin{figure}
\centering
\includegraphics[width=0.5\textwidth]{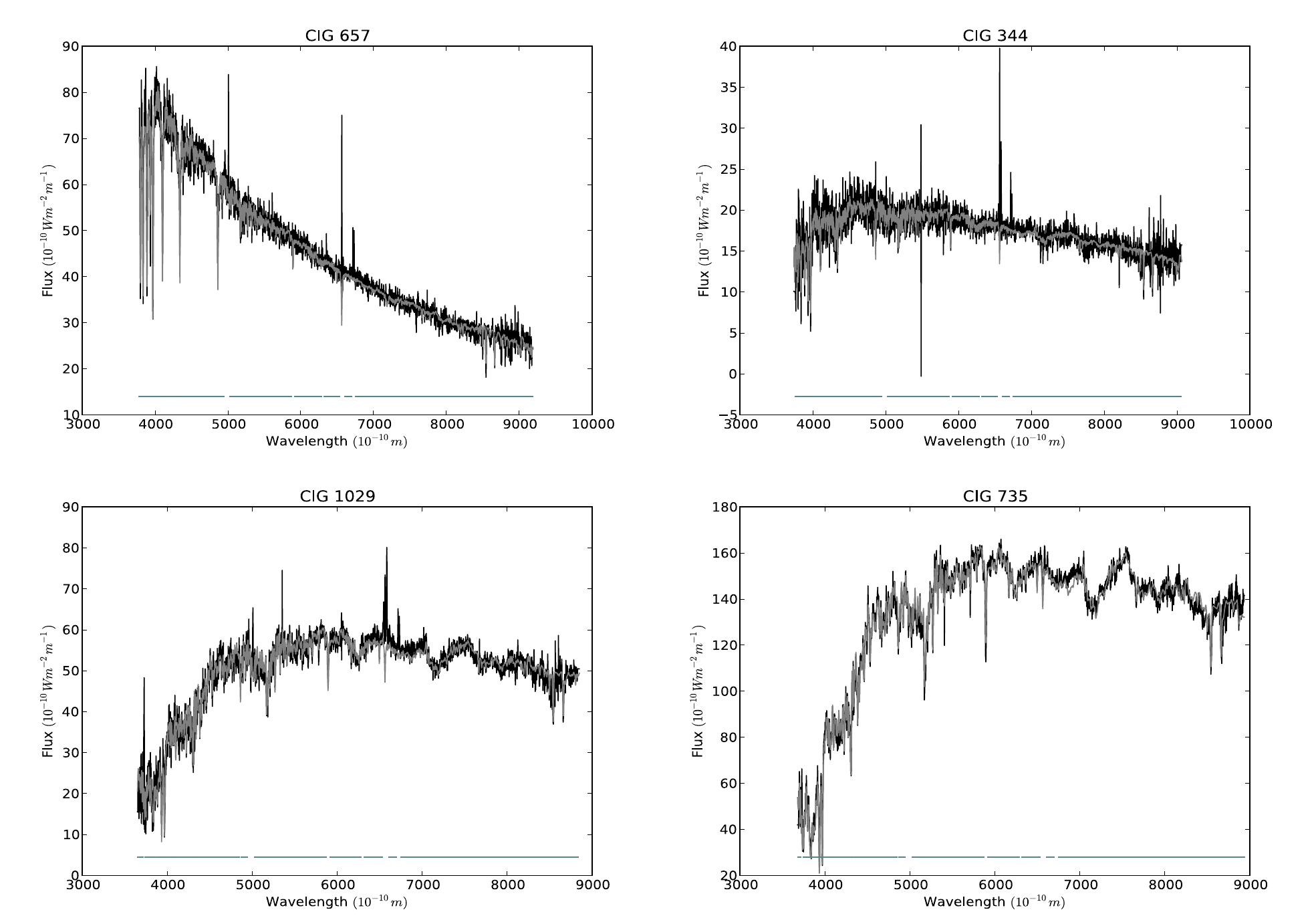}
\caption{Example of the fit of the underlying stellar populations performed by
Starlight for four AMIGA galaxies with different colours. 
The original spectrum is shown as a black line, the fitted
spectrum as a grey line and the line at the bottom indicates the
unmasked region at which the fit was performed. The average ages (log)
and metallicities of the estimated stellar populations for the depicted galaxies
are CIG 657 - 8.8 [a] - 0.005 Z; 
CIG 344 - 9.4 [a] - 0.015 Z; 
CIG 1029 - 9.9 [a] - 0.023 Z; 
CIG 735 - 10.1 [a] - 0.020 Z.}
\label{fig:fit_sl}
\end{figure}

\label{ref:av}
In Table~\ref{table:sl45} we present the following physical parameters estimated
by Starlight: the velocity shift of the spectrum with respect to the rest frame,
the velocity dispersion of the stellar populations, and the extinction. In a few
cases (4.7\%) Starlight found slightly negative extinctions, which may indicate
that the determination of the underlying stellar populations is not fully
reliable; however, the fitted spectra are sufficient to remove the absorption
features. All fits are clear-cut except for some of the few ($n \lesssim 7$)
galaxies that present strong broad lines. But the emission spectra of these
galaxies are not needed for their classification. In Table~\ref{table:sl45_pop}
we present the breakdown of stellar populations found by Starlight to best fit
each spectrum, indicated by the percentage of each different stellar population.
Four examples of fits are shown in Fig.~\ref{fig:fit_sl}.

\begin{table}
\caption{Starlight output.$^1$}
\label{table:sl45}
~\\
\centering
\begin{tabular}{r c c c c c}
\hline
CIG & v shift & v disp. & A$_{\nu}$ $^2$ \\
    &   km/s & km/s &   \\
\hline \hline
11&-36.82&55.14&0.6932\\
16&-37.38&93.08&-0.0750\\
19&-37.82&165.12&0.1241\\
56&3.03&114.42&0.6307\\
60&-16.99&52.00&0.5313\\
187&-46.87&79.07&0.3914\\
189&-39.79&183.45&0.0660\\
 ... & ... & ... & ...  \\
\hline
\end{tabular}
\begin{list}{}{}
 \item $^1$ \textit{SDSS sample}, narrow lines ($n=344$). Columns: 
(1) CIG catalogue number;
(2) shift in velocity computed by Starlight;
(3) velocity dispersion of the stellar component;
(4) extinction computed by Starlight.
 \item $^2$ In a few cases the computed extinction has slightly negative 
values (see text, Sect.~\ref{ref:av}).
\end{list}
\end{table}

\begin{table}
\caption{Starlight stellar populations.$^1$}
\label{table:sl45_pop}
~\\
\centering
\begin{tabular}{r r c c}
\hline
CIG & Age & Metallicity  & Percentage \\
    &  Ma & Z & \% \\
\hline \hline
11&1&0.004&0.0000\\
11&3.16&0.004&0.0000\\
11&5.01&0.004&2.9137\\
11&10&0.004&6.0817\\
 ... & ... & ... & ... \\
11&11000&0.050&1.9735\\
11&13000&0.050&0.0000\\
16&1.&0.004&0.0000\\
 ... & ... & ... & ... \\
\hline
\end{tabular}
\begin{list}{}{}
 \item $^1$ \textit{SDSS sample}, narrow lines, 45 populations per
galaxy
($n=344 \times 45$). Columns: 
(1) CIG catalogue number;
(2) age of the stellar population in megayears (Ma);
(3) metallicity;
(4) contribution of this stellar population.
\end{list}
\end{table}

The synthetic spectra obtained from the combination of stellar populations (as
indicated in Table~\ref{table:sl45_pop}) are available online at the AMIGA web
page (http://amiga.iaa.es/). 

\subsection{Emission lines}

Using the stellar populations derived from the Starlight
program, we estimated the separate contribution of the emission from the
interstellar medium, and the AGN emission, if present. The synthetic
spectrum of the stellar populations was subtracted from the initial spectrum for
each galaxy not flagged in Table~\ref{table:spec} (galaxies with narrow lines
and good data; $n = 344$), and the narrow lines found in the resulting spectra
were measured using a semi-automatic Gaussian fit.

As will be explained in the next section, there are specific emission lines that
can be used to provide a diagnostic test for the presence of an AGN. We
therefore performed the fitting in five different spectral regions: a) the
H$\beta$ region, b) the [\ion{O}{iii}] region, c) the [\ion{O}{i}] region, d)
the H$\alpha$ and [\ion{N}{ii}] region, and e) the [\ion{S}{ii}] region. If
nearby lines existed (closer than 0.0030 in $\Delta[\log_{10}(\lambda)]$ units),
we used a single fitting region for all nearby lines. Those regions correspond
to [\ion{N}{ii}] and H$\alpha$ and to [\ion{S}{ii}]. To remove any residual
continuum emission from the subtraction of the underlying stellar spectrum, we
subtracted a linear continuum baseline for each region. An estimate of the best
fit to the lines in each region was computed by developing an automatic
pipeline, then the estimated fit was visually checked and changed where needed.
Line fluxes were calculated using the area below the fitted Gaussian:
$\sqrt{2\pi}\left( \mathtt{height} \times \mathtt{width}\right) $, where
\texttt{width} is the standard deviation of the fitted Gaussian. The associated
error is proportional to the noise in the spectral region and is calculated as,
$\sqrt{2\pi}\left( \sigma \times \mathtt{width}\right) $, where $\sigma$ is the
rms continuum noise. When the line peak is below $3\sigma$ significance, the
upper limit is calculated as, $\sqrt{2\pi}\left( 3\sigma \times
\mathtt{width}\right) $. The line parameters are presented in
Table~\ref{table:lines}. The fluxes of the detected emission lines were
subsequently compared with those obtained from the MPA-JHU SDSS
catalogues\footnote{http://www.mpa-garching.mpg.de/SDSS/DR7/}. The line flux
ratios are consistent within the error for all measurements.

\begin{table*}
\caption{Measurement of the lines.$^{1,2}$}
\label{table:lines}
~\\
\centering
\begin{tabular}{r c c c c c c c}
\hline
CIG & Line & Flux  &  Peak intensity & Width & Pos. & $\sigma$ & Comment
\\
    & $10^{-10} \mathrm{m}$  
    & $10^{-20} \mathrm{W} \mathrm{m}^{-2}$ 
    & $10^{-10} \mathrm{W} \mathrm{m}^{-2}\,\mathrm{m}^{-1}$  
    & $10^{-10} \mathrm{m}$ 
    & $10^{-10} \mathrm{m}$ 
    & $10^{-10} \mathrm{W} \mathrm{m}^{-2}\,\mathrm{m}^{-1}$
    &  \\
\hline \hline
11 & H$\alpha$ (6563) & $ 566.4 \pm 3.7$ &136.1487&1.6597&6563.50&0.8954&   \\
11 & $\mathrm{[\ion{N}{ii}]}$ 6548 & $ 56.5 \pm 3.8$
&13.2799&1.6964&6548.78&0.8954&   \\
11 & $\mathrm{[\ion{N}{ii}]}$ 6583 & $ 186.9 \pm 3.8$
&43.9571&1.6964&6584.18&0.8954&   \\
 ... & ... & ... & ... & ... & ... & ... & ... \\
11 & $\mathrm{[\ion{S}{ii}]}$ 6730 & $ 61.5 \pm 3.6$
&14.7222&1.6671&6731.64&0.8616&   \\
16 & H$\alpha$ (6563) & $ 37.3 \pm 7.6$ &5.9584&2.5002&6560.72&1.2170&   \\
16 & $\mathrm{[\ion{N}{ii}]}$ 6548 & $< 22.9 \pm 7.6$
&1.9993&2.5002&6550.32&1.2170&   \\
 ... & ... & ... & ... & ... & ... & ... & ... \\
\hline
\end{tabular}
\begin{list}{}{}
\item[$^{1}$] \textit{SDSS sample}, narrow lines, eight lines per galaxy
($n=344 \times 8$). Columns:
(1) CIG catalogue number;
(2) name of the line. Numerical codes in the electronic version: 
 0 - H$\alpha$ ($6563 \times 10^{-10}$ m), 
  1 - $\mathrm{[\ion{N}{ii}]}$ $6548 \times 10^{-10}$ m,
  2 - $\mathrm{[\ion{N}{ii}]}$ $6583 \times 10^{-10}$ m,
  3 - H$\beta$ ($4861 \times 10^{-10}$ m),
  4 - $\mathrm{[\ion{O}{iii}]}$ $5007 \times 10^{-10}$ m,
  5 - $\mathrm{[\ion{O}{i}]}$ $6364 \times 10^{-10}$ m,
  6 - $\mathrm{[\ion{S}{ii}]}$ $6716 \times 10^{-10}$ m,
  7 - $\mathrm{[\ion{S}{ii}]}$ $6730 \times 10^{-10}$ m;
(3) flux of the line, the symbol '$<$' indicates that the flux is an upper
limit;
(4) peak intensity of the line;
(5) width of the line;
(6) central wavelength of the line;
(7) estimated noise for the line region;
(8) comments about the measurement of the line.
\item[$^{2}$] Equivalences of units: $10^{-20} \mathrm{W} \mathrm{m}^{-2}
\equiv 10^{-17} \mathrm{erg}\, \mathrm{s}^{-1} \mathrm{cm}^{-2}$ and $10^{-10}
\mathrm{W} \mathrm{m}^{-2}\,\mathrm{m}^{-1} \equiv 10^{-17} \mathrm{erg}\,
\mathrm{s}^{-1} \mathrm{cm}^{-2}\,\mathrm{\AA }^{-1}$.
\end{list}
\end{table*}

\section{Classification and properties of the AGN}
\label{sec:class}

\label{sec:broadlines}
Some galaxies in the SDSS sample show broad emission lines and were directly
classified as Seyfert 1. Since the Seyfert 1/Seyfert 2 separation was not the
main aim of the work, we classified the Seyfert type of the galaxies by visual
inspection. We found nine Seyfert 1 galaxies (CIG~204, 214, 336, 349, 719, 747,
749, 893, 1008) \footnote{AGN classification obtained from the literature in
\citet{Sabater2008}: CIG~214 - Sy 1.0; CIG~349 - Sy 1.5; CIG~719 - Sy 1.0;
CIG~1008 Sy 1.2} and among them seven clearly exhibited a broad H$\alpha$ line.
CIG~336 and 893 showed a faint broad H$\alpha$ component but, because of the
broad width of the [\ion{N}{ii}]\ lines we classified them as Seyfert 1. In a
future work we will use a parametric classification scheme like the one in
\citet{Hao2005} and \citet{Martinez2008}.

Galaxies presenting only narrow lines were classified using the standard
emission line diagnostic diagrams \citep{Baldwin1981,Veilleux1987}. Line ratios
adopted as suitable diagnostics  were a)
$\mathrm{log([\ion{O}{iii}]/\mathrm{H}\beta)}$, primarily an indicator of mean
ionization level and temperature; b)
$\mathrm{log([\ion{N}{ii}]/\mathrm{H}\alpha)}$, which is less immediately
obvious but provides a good separation between star-forming nuclei and AGN
\citep{Osterbrock2006}; c) $\mathrm{log([\ion{S}{ii}]/\mathrm{H}\alpha)}$ and
$\mathrm{log([\ion{O}{i}]/\mathrm{H}\alpha)}$, as indicators of the relative
importance of an extended partially ionised zone produced by high-energy
photoionization. These ratios were grouped to form the three following
diagnostic diagrams: 
\begin{itemize} \item
$\mathrm{log([\ion{O}{iii}]/\mathrm{H}\beta)}$ vs.
$\mathrm{log([\ion{N}{ii}]/\mathrm{H}\alpha)}$ (from here-on [\ion{N}{ii}]
diagram). \item $\mathrm{log([\ion{O}{iii}]/\mathrm{H}\beta)}$ vs.
$\mathrm{log([\ion{S}{ii}]/\mathrm{H}\alpha)}$ ([\ion{S}{ii}] diagram). \item
$\mathrm{log([\ion{O}{iii}]/\mathrm{H}\beta)}$ vs.
$\mathrm{log([\ion{O}{i}]/\mathrm{H}\alpha)}$ ([\ion{O}{i}] diagram).
\end{itemize} 
The line ratio measures used in the diagnostic diagrams for the
AMIGA SDSS sample are presented in Table~\ref{table:diag}. A ratio is given as
an upper or a lower limit if only an upper limit was measured for one of the
lines.  When both lines in a ratio involved upper limits, the corresponding
field is left blank.

\begin{table*}
\caption{Line ratios used for the diagnostic.$^1$}
\label{table:diag}
~\\
\centering
\begin{tabular}{r c c c c}
\hline
CIG & $\mathrm{log([\ion{N}{ii}]/\mathrm{H}\alpha)}$ & $\mathrm{log([\ion{O}{iii}]/\mathrm{H}\beta)}$ & $\mathrm{log([\ion{S}{ii}]/\mathrm{H}\alpha)}$ & $\mathrm{log([\ion{O}{i}]/\mathrm{H}\alpha)}$ \\
\hline \hline
11& $ -0.4815 \pm 0.0028$ & $ -0.614 \pm 0.015$ & $ -0.5814 \pm 0.0028$ & $ -1.6132 \pm 0.0028$ \\
16& $< -0.212 \pm 0.089$ & - & $< 0.035 \pm 0.089$ & $< -0.053 \pm 0.089$ \\
19& $ 0.025 \pm 0.067$ & $ 0.07 \pm 0.13$ & $ 0.011 \pm 0.067$ & $< -0.217 \pm 0.067$ \\
56& $ -0.31933 \pm 0.00090$ & $ -0.6795 \pm 0.0054$ & $ -0.63006 \pm 0.00090$ & $ -1.64935 \pm 0.00090$ \\
60& $ -0.4241 \pm 0.0027$ & $ -0.372 \pm 0.012$ & $ -0.3988 \pm 0.0027$ & $ -1.5158 \pm 0.0027$ \\
187& $ 0.128 \pm 0.027$ & $ 0.666 \pm 0.093$ & $ -0.278 \pm 0.027$ & $< -0.767 \pm 0.027$ \\
189& - & - & - & - \\
 ... & ... & ... & ... & ... \\
\hline
\end{tabular}
\begin{list}{}{}
\item[$^{1}$] \textit{SDSS sample}, narrow lines ($n=344$). The first
column is the CIG catalogue number. The other
columns are the different line ratios used in the diagnostic diagrams. The
presence of an upper or a lower limit is indicated with a '$<$' or a '$>$'
symbol, respectively. The precision of the numbers is adjusted for the error to
have two significant digits.
\end{list}
\end{table*}

\citet{Veilleux1987} proposed several semi-empirical boundaries within the
diagnostic diagrams including a theoretical starburst region and objects showing
other types of excitation such as AGN. LINERs were defined as a separate class
of nuclei by \citet{Heckman1980}, and \citet{Ho1997} defined similar
classification criteria using the [\ion{O}{i}] diagram to distinguish between
Seyfert 2 and LINER nuclei \citep[see also the review of][]{Ho2008}. We used the
theoretical curves proposed by \citet{Kewley2001} to separate narrow line AGN
(NLAGN; above the curves) from SF nuclei (SFN; below the curves; in the
[\ion{S}{ii}] and [\ion{O}{i}] diagrams) and the empirical curve of
\citet{Kauffmann2003} to select SF nuclei (SFN; below the curve in the
[\ion{N}{ii}] diagram). Galaxies between both curves in the [\ion{N}{ii}]
diagram were considered to harbour a transition object (TO; composite of SFN and
NLAGN). To separate Seyfert 2 and LINERs, we used the empirical classification
scheme provided by \citet{Kewley2006} for the [\ion{S}{ii}] and the [\ion{O}{i}]
diagrams. If only one of the two line ratios was available for a given
diagnostic diagram, 
a definitive classification can be obtained
from that line ratio in some cases \citep{Miller2003,Martinez2008}, e. g.
$\mathrm{log}([\ion{O}{iii}]/\mathrm{H}\beta) \geqslant 0.8$ or
$\mathrm{log}([\ion{N}{ii}]/\mathrm{H}\alpha) \geqslant 0.0$ for AGN.

Censored data (upper and lower limits) were taken into account for the
classification. Some galaxies can be unambiguously classified despite the
presence of an upper limit. For example, a galaxy located below the
\citet{Kauffmann2003} curve (SFN region) is classified as SFN even if there is
an upper limit in the $\mathrm{log}([\ion{N}{ii}]/\mathrm{H}\alpha)$ ratio. We
can classify 11\% more galaxies (increasing the overall classification fraction
from 84\% to 95\%) if upper limits are considered (see Sect.~\ref{sec:robust}
for a more detailed comparison).

\begin{figure*}
\centering
 \includegraphics[width=\textwidth]{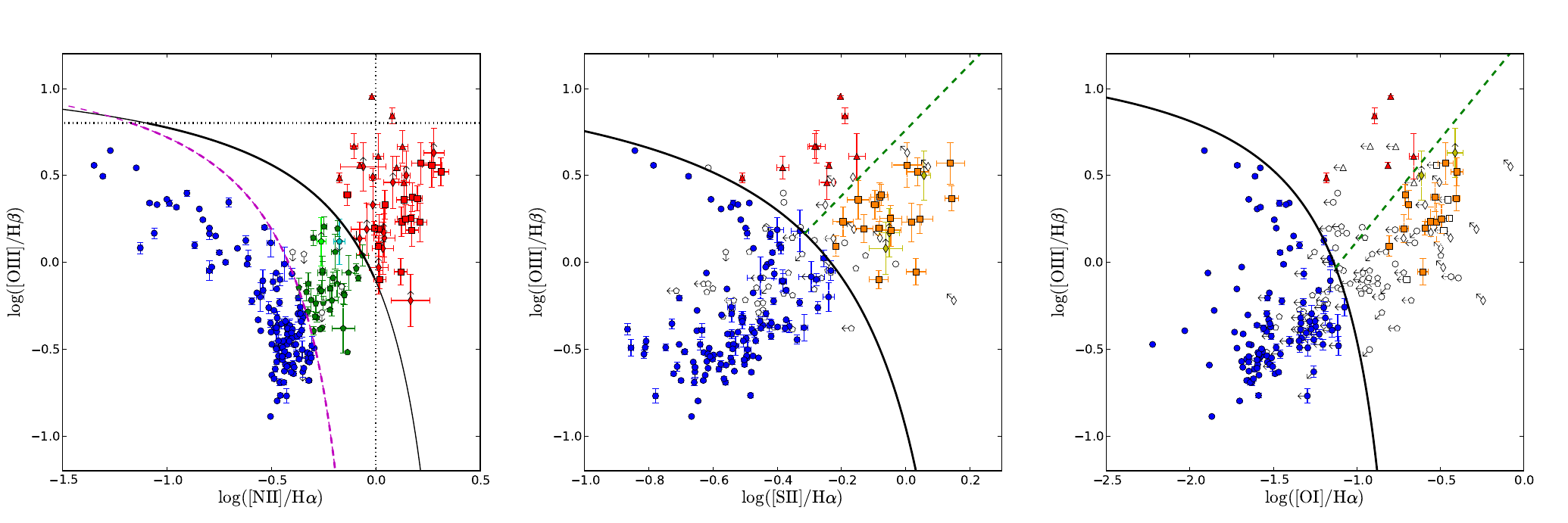}
   \caption{Emission line diagnostic diagrams. Error bars have been plotted, and
an arrow added for upper or lower limits. Left panel: [\ion{N}{ii}] diagnostic
diagram. The shape of the marker indicates the final classification of nuclear
activity: circle - SFN; pentagon - TO; triangle - Sy2; square - LINER, diamond -
NLAGN. Filled markers correspond to galaxies classified using this diagnostic
diagram, and the colour indicates the derived classification: red - AGN; green -
TO; blue - SFN; cyan - TO or SFN; light green - TO or AGN. See Section
\ref{sec:class} for a complete description of the classification scheme. Middle
panel: [\ion{S}{ii}] diagnostic diagram. The AGN-SFN separation curve of
\citet{Kewley2001} is shown as a black solid line and the Sy2-LINER separation
line of \citet{Kewley2006} as a green dashed line. Symbol shapes are the same as
in the [\ion{N}{ii}] diagram and the colours indicate the classification
obtained with this diagram: red - Sy2; yellow - NLAGN; orange - LINER; and blue
- SFN. Right panel: [\ion{O}{i}] diagnostic diagram. Symbol shapes are the same
as in the [\ion{N}{ii}] diagram. Colours and separation curves are the same as
in the [\ion{S}{ii}] diagram. Empty symbols correspond to galaxies that cannot
be unambiguously classified using this diagram but that were eventually
classified using the other diagrams (see text).
           }
      \label{fig:diag}
\end{figure*}

The left panel of Fig.~\ref{fig:diag} shows the [\ion{N}{ii}] diagnostic diagram
for the SDSS complete sample (although classification criteria were applied to
all galaxies in the SDSS sample). Galaxies classified as NLAGN typically have
higher errors in their emission line ratios because the flux of their emission
lines is usually lower than for SFN types. In the central and right panels of
Fig.~\ref{fig:diag} the [\ion{S}{ii}] and [\ion{O}{i}] diagnostic diagrams are
shown. The transition objects are widely spread in both the NLAGN and
star-forming regions, as is usually found in the literature.

To adopt a final classification for each galaxy, we proceeded as follows. The
classification obtained from the [\ion{N}{ii}] diagram was given precedence over
classifications inferred from other diagrams. In most cases there was agreement 
(although there are nine cases where a galaxy is classified as SFN using
[\ion{N}{ii}] and as AGN based on the other diagrams or vice versa). If an
object could not be unambiguously classified using this diagram, the
classification obtained with the other two diagrams was used if the two
classifications agreed. We could not distinguish between TO or NLAGN for three
galaxies, and between TO or SFN in three other cases. In these cases, we added 
both classifications to the database. Notice that some of the NLAGN cannot be
sub-classified as LINER or Seyfert 2 using the [\ion{S}{ii}] and [\ion{O}{i}]
diagrams and were entered as NLAGN in the database. If a transition object was
classified as an AGN or SFN in the [\ion{S}{ii}] and [\ion{O}{i}] diagrams, we
indicate this classification in the database in an additional ``transition
object type'' column. The final classification is presented in
Table~\ref{table:class}, in which galaxies that presented broad lines have been
added as Seyfert 1.

\begin{table}
\caption{Classification.$^{1,2}$}
\label{table:class}
~\\
\centering
\begin{tabular}{r c c c c c}
\hline
CIG & [\ion{N}{ii}] & [\ion{S}{ii}] & [\ion{O}{i}] & \textbf{Final}  & TO \\
    & class.        & class.        & class.       & \textbf{class.} &
type$^{3}$
\\
\hline \hline
11&       SFN&       SFN&       SFN&       SFN&    -\\
16&        -&        -&        -&        -&    -\\
19&    NLAGN&    LINER&        -&    LINER&    -\\
56&       SFN&       SFN&       SFN&       SFN&    -\\
60&       SFN&       SFN&       SFN&       SFN&    -\\
187&    NLAGN&      Sy2&        -&      Sy2&    -\\
 ... & ... & ... & ... & ... & ... \\
228&       TO&       SFN&       SFN&       TO&   SFN\\
 ... & ... & ... & ... & ... & ... \\
\hline
\end{tabular}
\begin{list}{}{}
\item[$^{1}$] \textit{SDSS sample} ($n=353$). Columns: 
(1) CIG number; 
(2) classification with the [\ion{N}{ii}] diagnostic diagram; 
(3) classification with the [\ion{S}{ii}] diagnostic diagram;
(4) classification with the [\ion{O}{i}] diagnostic diagram;
(5) final classification adopted for our study;
(6) transition object subtype.
\item[$^{2}$] Acronyms:
SFN - star-forming nucleus;
TO - transition object;
NLAGN - narrow line AGN (Seyfert 2 or LINER);
LINER - LINER;
Sy1 - Seyfert 1;
Sy2 - Seyfert 2;
UNK - unknown (only for the TO subtype). 
Three galaxies have no unambiguous classification between TO and NLAGN (``TO or
NLAGN'') and three more between TO and SFN (``TO or SFN''), as explained in the
text. 
\item[$^{3}$] Subtype of transition object if classified as a TO 
(see text).
\end{list}
\end{table}

The definition of the lines of separation between different types of galactic
nuclei in the diagnostic diagrams can affect the final classification of a
galaxy (e.g. see \citealt{Kewley2006} and \citealt{Constantin2006}). Differences
in the classification also arise from the method chosen for the final
classification from the different diagnostics diagrams where these disagree, or
the different handling of non-detected or upper limits in the line flux. New
alternative classification methods \citep{Buttiglione2010,CidFernandes2010} use
different combinations of lines or derived parameters and classification
schemes. \citet{Stasinska2008} studied the possibility that the emission of some
of the galaxies in the region of AGN could actually be powered by stellar
processes \citep[hot post-asymptotic giant branch stars and white
dwarfs;][]{Binette1994}. The classification proposed by \citet{CidFernandes2010}
allows one to separate between ``retired galaxies'' (powered by these processes)
and weak accretion-powered AGN \citep{Cid-Fernandes2011}. All these differences
in the definitions must therefore be taken into account to compare between
different samples. For this reason, to allow a fair unbiased comparison with
other samples, we kept and release all information (emission line fluxes,
detection information, emission line ratios, classification for each diagnostic
diagram and final classification) needed to allow a different classification
scheme to be applied.

Using our classification criteria, we were able to perform a statistical study
of the different types of galaxies found in the sample. The values obtained are
listed in Table~\ref{table:sdss_stat} both for the SDSS sample and the SDSS
complete sample, although only the latter ones were considered for statistical
purposes. The prevalence of optically selected active galactic nuclei for the
AMIGA sample of isolated galaxies is 20.4\%, or 36.7\% including TOs. 

\begin{table}
\caption{Statistics of nuclear activity in AMIGA galaxies.}
\label{table:sdss_stat}
~\\
\centering
\begin{tabular}{lrrr}
\hline
Classification & Total & Complete  & Percentage \\
               & sample & sample & (complete sample) \\
\hline
\hline
Unclassified:             &  27 &  11 &   4.9\% \\
~~Without emission lines  &   8 &   4 &   1.8\% \\
~~With emission line/s    &  19 &   7 &   3.1\% \\
\hline
Not unamb. classified:    &   5 &   2 &   0.9\% \\
~~TO or NLAGN             &   1 &   1 &   0.4\% \\
~~TO or SFN               &   4 &   1 &   0.4\% \\
\hline
Classified:               & 321 & 213 &  94.2\% \\
~~SFN                     & 191 & 130 &  57.5\% \\
~~TO                      &  53 &  37 &  16.4\% \\
~~AGN:                    &  77 &  46 &  20.4\% \\
~~~~NLAGN total:          &  68 &  41 &  18.1\% \\
~~~~~~LINER               &  27 &  18 &   8.0\% \\
~~~~~~Sy2                 &  10 &   9 &   4.0\% \\
~~~~~~NLAGN               &  31 &  14 &   6.2\% \\
~~~~Sy1                   &   9 &   5 &   2.2\% \\
~~TO + AGN:               & 130 &  83 &  36.7\% \\
\hline
Total:                    & 353 & 226 & 100.0\% \\
\hline
\end{tabular}
\end{table}

\subsection{Robustness of the classification method}
\label{sec:robust}

The fraction of galaxies that can be classified and their final classification 
may depend on factors such as the signal-to-noise level of the spectrum, the
classification method used, or the criteria used to assign a given type of
nuclear activity depending on the classification obtained from the diagnostic
diagrams.

\begin{figure}
   \centering
   \includegraphics[width=0.48\textwidth]{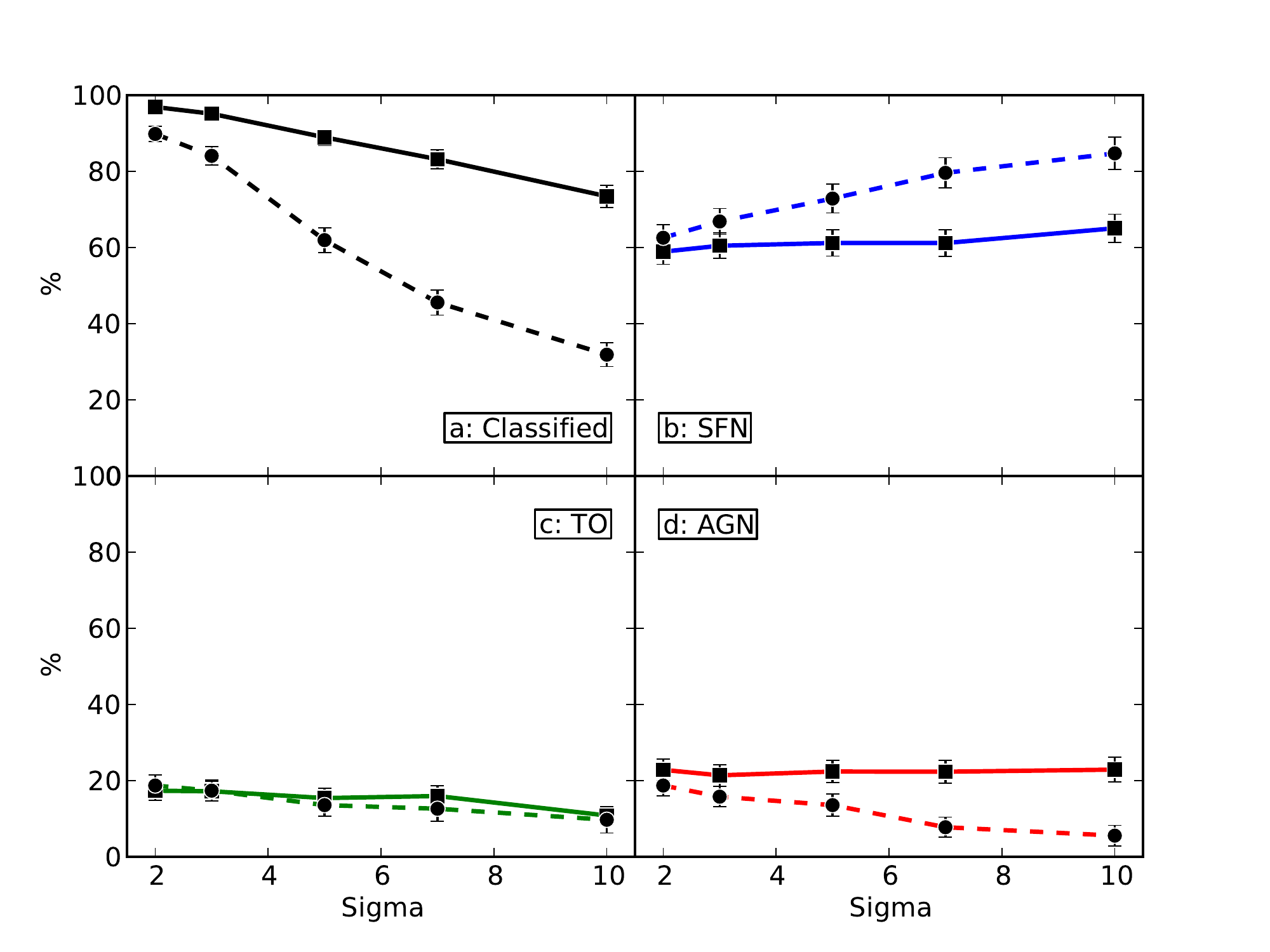}
   \caption{Fraction of galaxies that are a) classified; b)
classified as SFN; c) classified as TO; d) classified as AGN, with respect to
the noise level using two different
classification methods. The first method is the one presented in this paper that
takes into account the information provided by upper limits (solid line,
square markers) and the second method is the classical method that takes into
account only galaxies with detections of all lines involved in the
[\ion{N}{ii}] diagnostic diagram (dashed line, round markers).
           }
   \label{fig:sigma}
\end{figure}

We tested the robustness of our classification method with respect to the
signal-to-noise level of the spectra. The required detection level of the
emission lines was raised from $3\sigma$ to $5\sigma$, $7\sigma$ and $10\sigma$
(higher $\sigma$ cut levels mimic a $3\sigma$ selection in observations taken
with progresively lower signal-to-noise ratios). Then, we re-calculated the
fraction of galaxies of each type using a) our method, which takes into account
the information carried by upper limits and b) a classical classification
method, which requires the detection of the four main emission lines \citep[e.g.
][]{Kewley2006}. The numbers are presented in Figure~\ref{fig:sigma}. The
fraction of classified galaxies drops with the noise for both methods, but the
drop is steeper for the classical method and, while with a $10\sigma$ cut level
our method classifies $\sim$73\% of the galaxies, the classical method
classifies only $\sim$32\% of the galaxies. The fraction of different types of
galaxies remains almost constant with respect to the noise using our method,
with a slight increase of the fraction of SFN with high cut levels ($10\sigma$)
probably originating from the slight decrease of TOs. The classical method tends
to overestimate the fraction of SFN and underestimate the fraction of AGN as the
noise increases because only SFN galaxies have sufficiently strong emission
lines to be detected over the noise. These figures suggest that our method not
only obtains a higher rate of classifications, but also tends to be more robust
in the classification of different types of galactic nuclei with respect to the
signal-to-noise level. 

We also checked the effect of lowering the detection level to $2\sigma$ to
include weak emission lines \citep{CidFernandes2010}. We risked considering as
detections some noise artefacts, but we can check if we are excluding many weak
active nuclei. The overall effect for our sample is a very slight increase in
the fraction of AGN galaxies. The relative fraction of different types of AGN
over the total number of AGN is similar at 2, 3 and $5\sigma$. From $7\sigma$
on, there is a drop in the fraction of LINERs ($\approx 70\%$ of the classified
NLAGN at 2, 3 and $5\sigma$; $\approx 38\%$ at $7\sigma$ and 20\% at $10\sigma$)
probably caused by missing retired galaxies with weak emission lines
\citep{Cid-Fernandes2011}.

\subsection{Properties of the host galaxies}

The relation between the chance for a galaxy to harbour an AGN and its stellar
mass has been well-established for different manifestations of nuclear activity:
optical \citep[e.g., ][]{Kauffmann2003}, radio \citep{Best2005b}, or X-ray
\citep{Tasse2011}. The larger the stellar mass, the higher the probability of
harbouring an AGN. It is also well known that there is a relation between
nuclear activity and morphology \citep{Moles1995}. Accordingly, a higher
fraction of AGN would be expected in more luminous galaxies and in earlier
morphological types.

\begin{figure}
\centering
\includegraphics[width=0.5\textwidth]{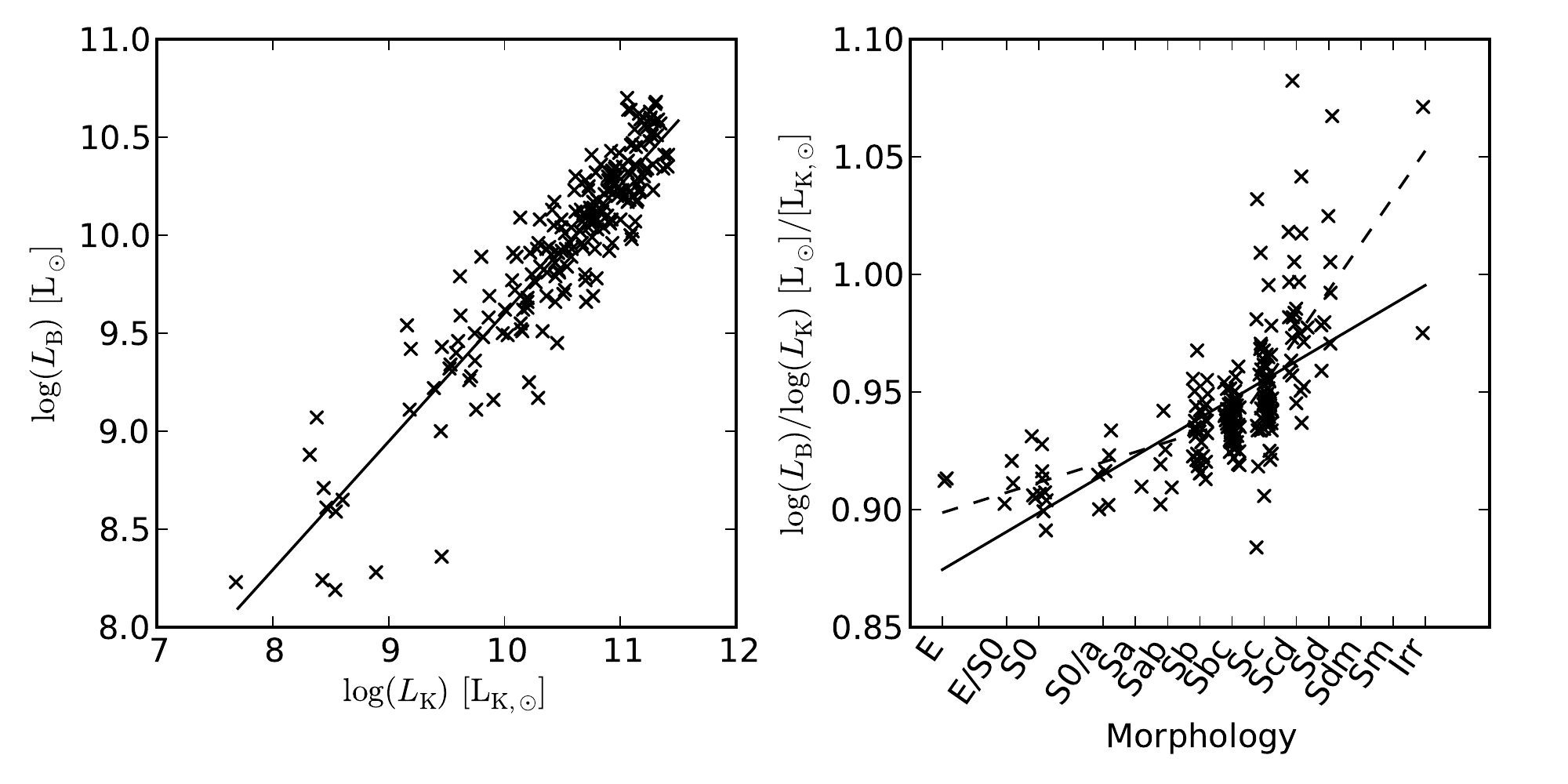}
\caption{Luminosities of the AMIGA isolated galaxies that are in
the SDSS complete sample and have available NIR luminosities (n = 211). The
left panel shows the relation between $L_{\rm{B}}$ and $L_{\rm{K}}$.
The ordinary least-squares (OLS) regression line is shown as a solid black line.
The right panel shows the relation of $\log(L_{\rm{B}})/\log(L_{\rm{K}})$ with
the morphology. The OLS regression for the whole sample is plotted as a solid
line. Two additional OLS regression lines are shown as dashed lines, the first
for galaxies with morphological types between E and Sbc and the second for
the rest. A small random scatter (from -0.25 to 0.25 of a class) was added
to the morphology values to allow a proper visualisation of the symbols.}
\label{fig:prop_lt}
\end{figure}

\begin{figure}
   \centering
   \includegraphics[width=0.5\textwidth]{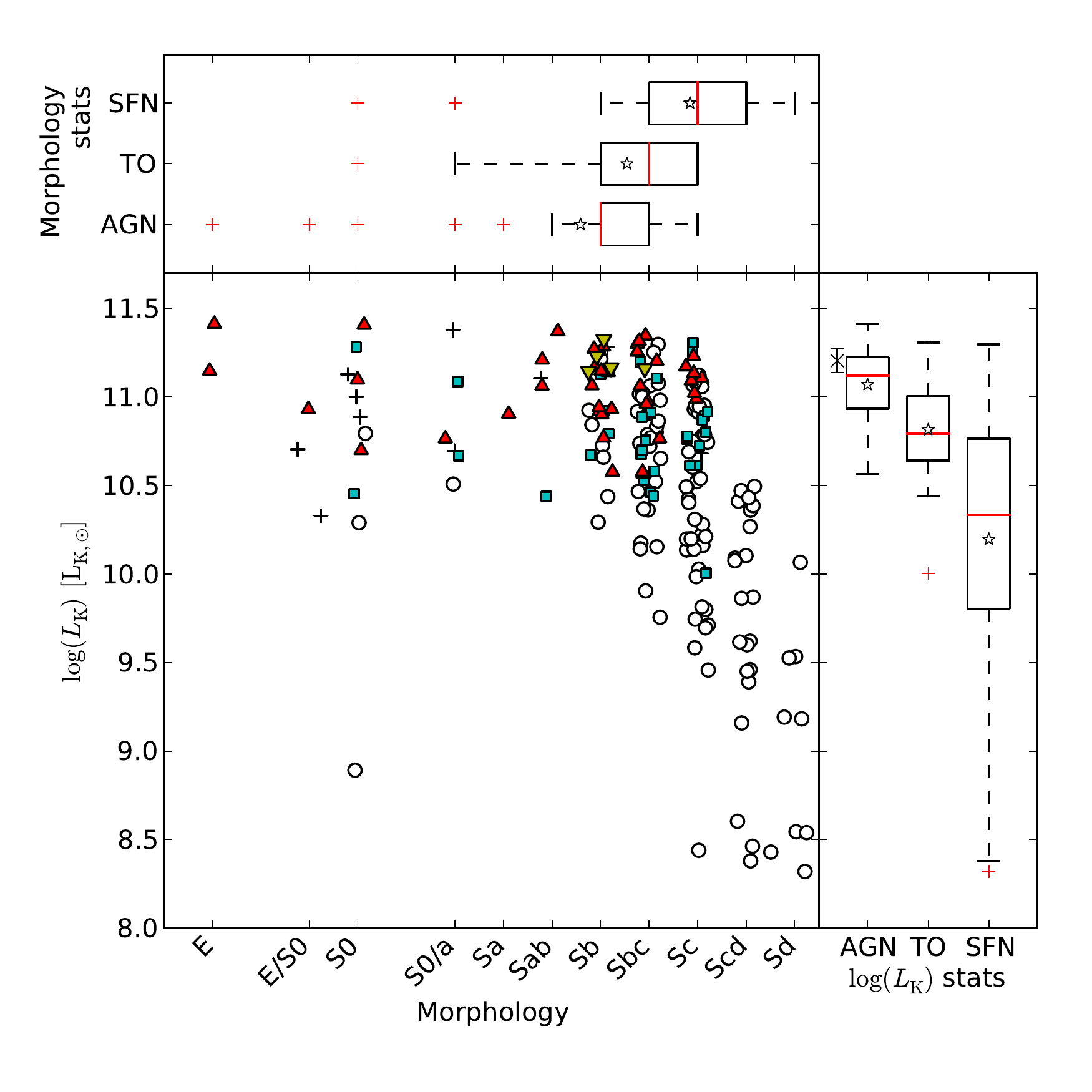}
   \caption{Properties of the AMIGA isolated galaxies that are in the SDSS
complete sample and have available NIR luminosities (n = 211). The main panel
shows $L_{\rm{K}}$ versus the morphology. The SFN are marked as white round
symbols, unclassified galaxies as black + symbols, TOs as cyan squares, and AGN
as triangles (red triangles in general and yellow inverted triangles for the
Seyfert 1). A small random scatter, like in Fig.~\ref{fig:prop_lt}, was added to
the morphology values. The additional panels show the statistics for different
types of nuclear activity with respect to the morphology (upper panel) and
$L_{\rm{K}}$ (right panel). The box marks the lower and upper quartiles, the
median is indicated as a red line inside the box, the whiskers extend to the
most extreme datapoint within $1.5 \times (x_{75\%}-x_{25\%})$ from the median,
and the outliers are marked as plus symbols. The mean values are marked using a
star symbol. The mean of $L_{\rm{K}}$ for Seyfert 1 galaxies is indicated in the
right panel with a cross at the left of the AGN box.
           }
   \label{fig:prop}
\end{figure}

We studied some properties of the host galaxies with respect to the different
types of nuclear activity found in the AMIGA sample. We considered only galaxies
belonging to the SDSS complete sample to ensure statistical robustness. The
relation of $L_{\rm{B}}$ with $L_{\rm{K}}$ is almost linear for AMIGA galaxies
(Fig.~\ref{fig:prop_lt}). The fraction $L_{\rm{B}}/L_{\rm{K}}$ has a mean value
of $0.946 \pm 0.030$ with a maximum value of 1.082 and a minimum of 0.884 and
presents a clear correlation with the morphology, with higher values for
late-types (Fig.~\ref{fig:prop_lt}). In Fig.~\ref{fig:prop} the relation of
$L_{\rm{K}}$ and the morphological type for galaxies in the SDSS complete sample
that have NIR fluxes available (n = 211; $\sim$93.4\%) is shown. We show the
statistics for AGN, TOs and SFN as well. Active galactic nuclei are hosted by
earlier types and higher luminosity galaxies than TOs, while just the opposite
happens for SFN. Because early types present a bigger fraction of the galaxy
mass in the spheroidal component and AMIGA galaxies harbour pseudo-bulges
instead of classic bulges \citep{Durbala2009}, there might be a relation between
the prevalence of AGN and the mass of the pseudo-bulge in AMIGA isolated
galaxies.

\begin{figure}
   \centering
   \includegraphics[width=0.5\textwidth]{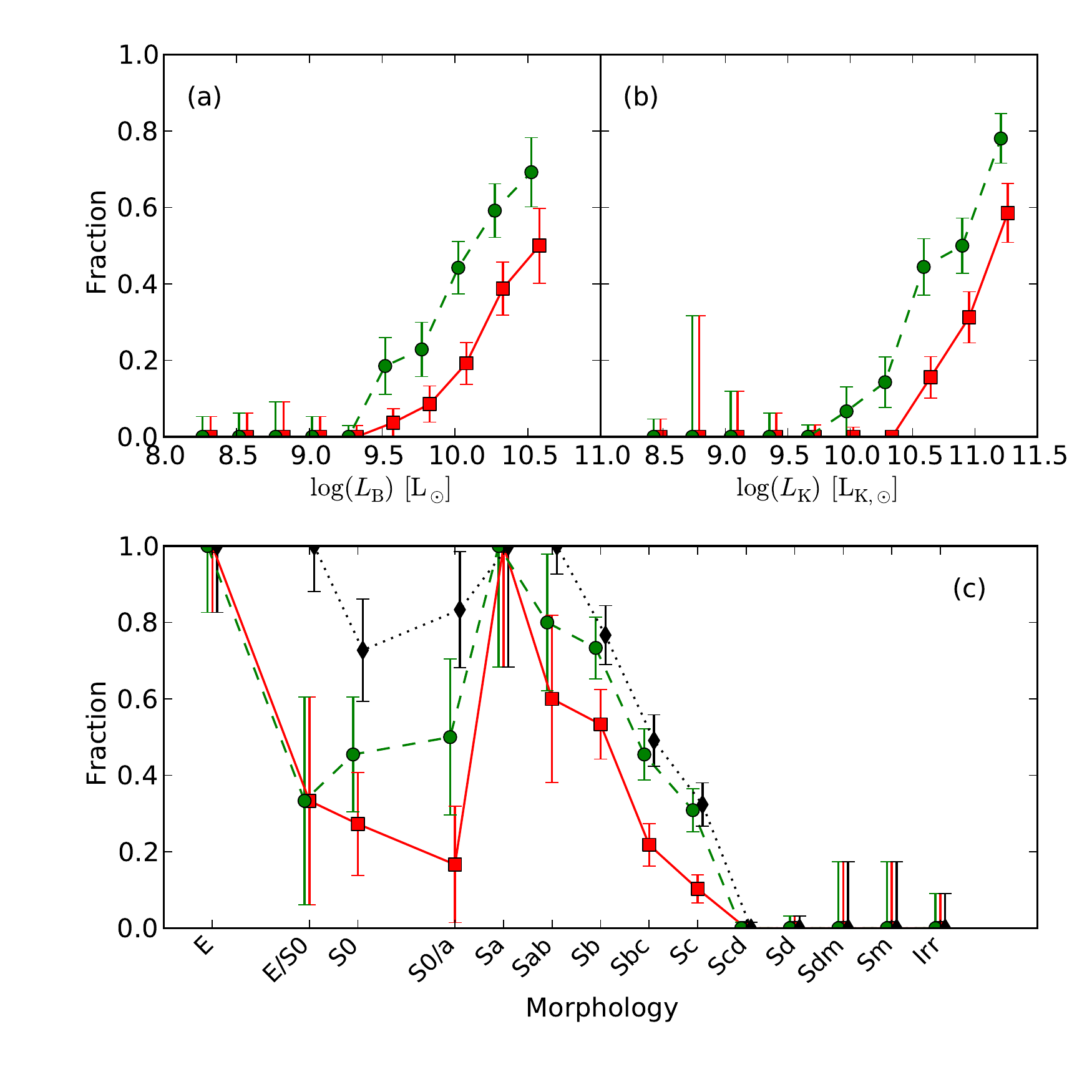}
   \caption{Relation between the fraction of AGN (red solid line, square
symbols) or AGN plus TOs (green dashed line, round symbols) with the optical
luminosity (a), NIR luminosity (b), and morphology (c) for the SDSS complete
sample. In the morphology panel we also show the fraction (black dotted line, 
diamond symbols) of
AGN plus TOs plus unclassified galaxies (galaxies without detected
emission lines that are often referred to in the literature as passive
galaxies).
           }
   \label{fig:frac}
\end{figure}

In Fig.~\ref{fig:frac} we show the fraction of AGN and AGN + TOs with respect to
optical luminosity ($L_{\rm{B}}$; panel a), $L_{\rm{K}}$ (proxy for stellar
mass; panel b) and morphology (panel c). We also considered AGN plus TOs plus
non-classified galaxies in the case of the morphology. There is a monotonic
increase of the fraction of AGN galaxies with respect to $L_{\rm{B}}$ and
$L_{\rm{K}}$, similar to the one found in the literature. The trend is also
similar considering the transition objects. In terms of morphology, there are no
AGN or TOs found in very late-type (Scd to Irregular) galaxies. There is a
monotonic increase in the fraction of AGN (from 0\% to 100\%) from type Sc to
Sa, while earlier-types yield noisier results because there are few galaxies in
these bins. There are four early-type galaxies presenting an SFN. Three of them
(CIG~364, 498 and 620) are edge-on galaxies classified as S0 or S0/Sa, and
CIG~503 is an S0 galaxy presenting Wolf-Rayet signatures \citep{Brinchmann2008}
and identified as an early-type star-forming galaxy by \citet{Temi2009} and
\citet{Wei2010}. The increase of the fraction of AGN is steeper with respect to
$L_{\rm{K}}$ than with respect to $L_{\rm{B}}$ due to the correlation between
$L_{\rm{B}}/L_{\rm{K}}$ and the morphological type: galaxies with higher
$L_{\rm{K}}$ and later type tend to have lower $L_{\rm{B}}$ and vice versa,
thus, reducing the steepness of the slope (see Fig.~\ref{fig:prop_lt}).

Seyfert 1 galaxies tend to present higher luminosities than narrow-line AGN
galaxies, as shown in Fig.~\ref{fig:prop}. This difference may be produced by
weak broad lines that are missed in galaxies with lower luminosities because of
the lower signal-to-noise level or the direct detection of the AGN light in the
luminosity measurements. There is no clear dependence on luminosity for the
other types of AGN, but the fraction of LINERs relative to the number of
classified NLAGN (LINER + Sy2) depends strongly on the morphological type. It
ranges from 25\% for Sc types to 83\% for Sab and earlier types (Sc - 25\%; Sbc
- 50\%; Sb - 82\%; Sab and earlier - 83\%). This trend is consistent with some
of the LINERs being actually retired galaxies ionized by low-mass evolved stars
\citep{Cid-Fernandes2011}.

\section{Comparison with galaxies in isolated denser environments}
\label{sec:comp}

One of the main aims of this study is to determine the effect of the environment
and interactions on the prevalence of nuclear activity. After estimating the
normal level of nuclear activity for a sample of galaxies not affected by major
interactions, a careful comparison with samples of galaxies in denser
environments is required to quantify the effect of the environment/interaction.
An unbiased comparison will require to take into account the effect of the
density-morphology and the density-luminosity relations
\citep{Dressler1980,Kauffmann2004,Blanton2005,Deng2009,Park2009, Deng2011}. The
strong correlation of the fraction of AGN with morphology and luminosity (both
optical and NIR) could otherwise produce a strong bias when comparing with
samples that could be composed of galaxies with different morphologies and
luminosities. 

Some of the densest environments in the universe are in compact groups of
galaxies, with projected galaxy densities similar to those found in the cores of
dense clusters. Among them, Hickson Compact Groups \citep[HCG; characterized by
4 to $\sim 10$ members with a low velocity dispersion $\sim 200$ km
s$^{−1}$;][]{Hickson1982} have been selected using an isolation criterion
\citep{Hickson1982,Sulentic1987,Hickson1992}. This combination of high galaxy
density in a low-density environment makes them a unique sample for comparison
with isolated galaxies by minimising the possible effects of the large-scale
environment. It is important to remark that AMIGA and HCG samples
were defined using only isolation criteria.

 \citet{Martinez2010} studied the optical nuclear activity of a complete
well-defined sample of HCG galaxies ($n = 270$) obtaining new spectroscopic
measurements ($n = 200$), as well as archival spectra ($n = 11$) and emission
line measurements from the literature ($n = 59$). The spectra were corrected for
absorption features from the underlying stellar populations, furthermore, the
authors followed a classification scheme very similar to the one that we present
in this paper: the diagnostic diagram used is the [\ion{N}{ii}] one, separation
lines are identical to ours and the
$\mathrm{log}([\ion{N}{ii}]/\mathrm{H}\alpha)$ line ratio is used when it is the
only one with confident values. The mean signal-to-noise ratios of the
$\mathrm{log}([\ion{N}{ii}]/\mathrm{H}\alpha)$ coefficient for both samples are
very similar (HCG - $\sim42$; AMIGA - $\sim56$).

The origin of the luminosity and morphology data used by \citet{Martinez2008} is
HyperLeda \citep{Paturel2003}. We also obtained luminosity and morphology data
from the HyperLeda catalogue for AMIGA galaxies. AMIGA curated data (presented
in Sect.~\ref{sec:lt}) present some fundamental differences compared to
Hyperleda data, which render a direct comparison between them unsuitable. In
particular, the optical luminosity is corrected for different effects
\citep{Mirian2012} and presents a systematic shift with respect to Hyperleda
data. The difference in morphological classification, although very low on
average ($-0.6$), presents a high dispersion ($2.5$). Hence, we used HyperLeda
data for the comparison between AMIGA and HCG to allow an homogeneous
comparison. The morphological classification is coded into numbers using the RC3
morphological types system and the absolute magnitude was transformed into
optical luminosity (B-band) using the relation $\log(L_{\rm{B}}) =
(5.51-M_{\rm{B}})/2.5\,[L_{\sun}]$. Finally, there are data for all but two
AMIGA galaxies of the SDSS complete sample ($n = 224$) and for all but one
galaxy of the HCG sample ($n=269$). The ranges and means of the optical
luminosity are very similar for both samples (AMIGA: $10.31 \pm 0.49$; HCG:
$10.4 \pm 2.0$) but the distribution of morphologies is quite different (AMIGA:
$3.6 \pm 2.5$; HCG: $0.4 \pm 3.6$) with earlier types in compact groups (see
upper panels of Fig.~\ref{fig:agn_comp}). We will present a method to take this
different distribution into account.

\subsection{Comparison method}
\label{sec:comp_method}

An empirical probability density function (\textit{pdf}) is built that
considers the optical luminosity, morphology, and the classification as an AGN
for a sample of galaxies. The definition is as follows:
$$
 f(L,t,a) = \frac{1}{n} \sum_{i=1:a=a_{i}}^{n}\frac{1}{2
\pi
\sigma_{L,i} \sigma_{t,i}} \\
\exp \left( \frac{(L -
L_{i})^{2}}{2 \sigma_{t,i}^{2}}\right)  \exp \left(\frac{(t -
t_{i})^{2}}{2 \sigma_{t,i}^{2}}\right),
$$
where $L$ is the luminosity, $t$ the morphology in the RC3 system, and $a$ is a
value that indicates whether the galaxy harbours an AGN ($a = 1$) or not ($a =
0$). Note that in the summation term, a galaxy $i$ is only added if
$a$ and $a_{i}$ are equal.

The partial probability density function with respect to the AGN activity
represents the distribution in luminosity and morphology of the sample and is
defined as $f(L,t) \equiv \sum_{a}^{} f(L,t,a) = f(L,t,1) + f(L,t,0)$. We can
obtain for each $L$ and $t$ pair the mathematical expectation with respect to
the values of \textit{a}: $$E[a](L,t) = f(L,t,1)/f(L,t) \equiv p_{a}(L,t),$$
Hence, the probability for a galaxy to harbour an AGN for a given luminosity
$L_{0}$ and morphology $t_{0}$, derived from our empirical \textit{pdf}, is:
$p_{a = 1}(L_{0},t_{0}).$
To compare the prevalence of AGN in two different samples, we can
compare the actual number of AGN galaxies in the reference sample with the
estimated number of AGN galaxies for a second sample calculated using the same
$p_{a = 1}(L,t)$ but weighted by the actual distribution of $L$ and $t$ in the
second sample. In other words, the
estimated number of AGN galaxies in a region (R) of the $L,t$ plane for the
second sample ($s2$), using the distribution of the first sample ($s1$), is
$$n_{s2,AGN} = n_{s2} \iint_{R} f_{s2}(L,t)p_{s1;a = 1}(L,t)
\mathrm{d}L
\mathrm{d}t.$$
In that way, we can check whether estimated and actual values are compatible
within the error (or not), in a given region, taking into account the effect of
the luminosity and morphology distribution. 

\subsection{Comparison with galaxies in compact groups}

\begin{figure}
   \centering
   \includegraphics[width=0.48\textwidth]{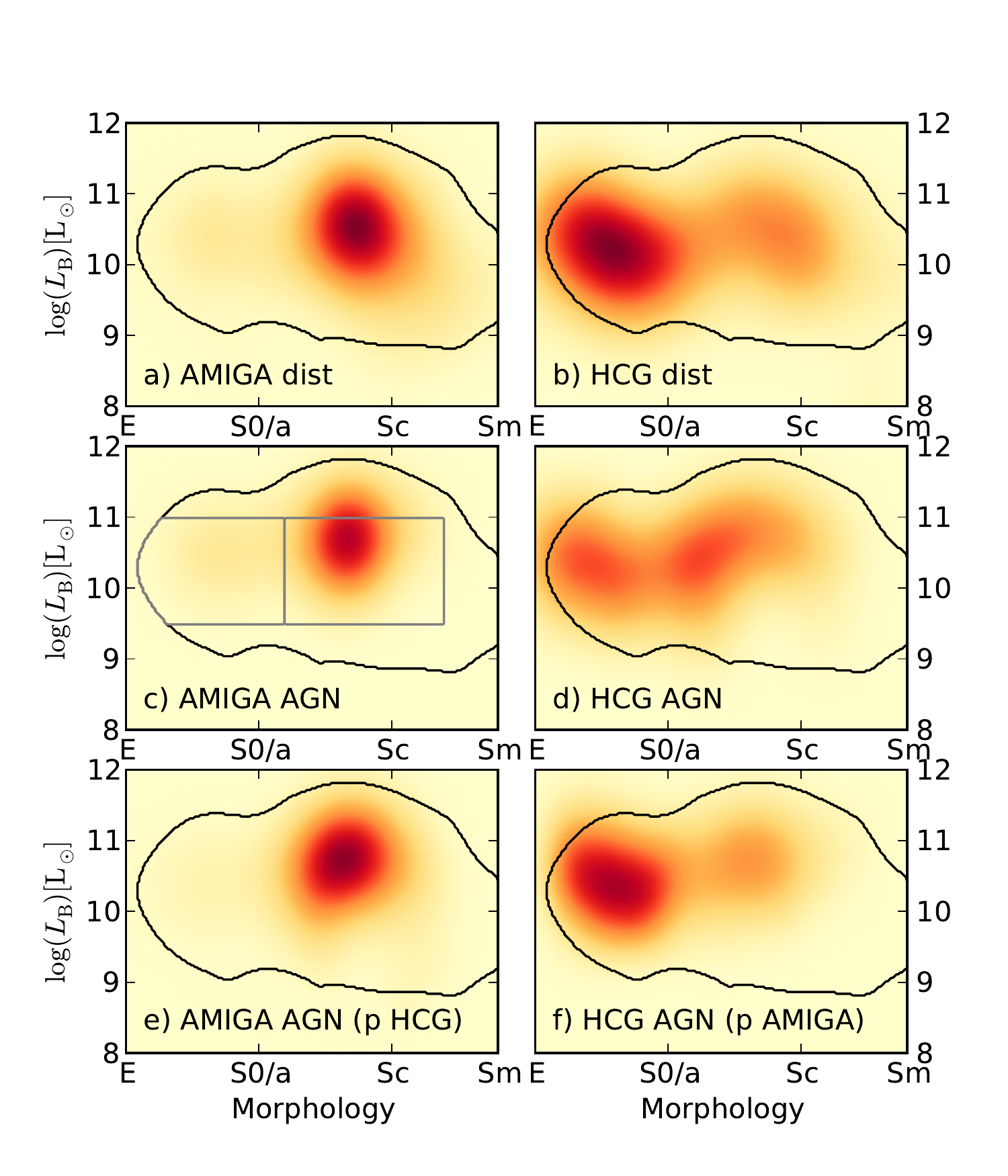}
   \caption{Distribution of galaxies (morphology and optical luminosity) and
comparison of the actual and estimated distribution of AGN in the AMIGA and HCG
samples. In the top panels the total distribution of galaxies for the AMIGA (a)
and HCG (b) samples are shown. The colour ranges from red to yellow, with red
corresponding to a higher density of galaxies. In the middle panels the
distributions of AGN in the AMIGA (c) and HCG (d) are shown. In the lower
panels, the cross-estimated distributions of AGN in AMIGA (e) and HCG (f) are
shown, as produced by combining the actual distribution of galaxies with the
probability of harbouring an AGN determined from the other sample. The colour
scale of the middle and lower panels is one third of the scale of the
corresponding upper panel. The black contour marks the region in common between
both samples (95.45\% of the galaxies inside the contour for each sample) and
the rectangles shown in the panel ``c'' denote the bins chosen for the
$\chi^{2}$ test.}
   \label{fig:agn_comp}
\end{figure}

There are 46 (20.5\%) galaxies classified as AGN in the AMIGA sample and 66
(24.5\%) galaxies classified as AGN in the HCG sample. The actual distribution
of galaxies and AGN are shown in the upper panels of Fig.~\ref{fig:agn_comp}.
Clearly, the distribution of galaxies is different in the two samples with later
types in the AMIGA sample and earlier types in the HCG one. The distribution of
AGN also looks different, but this may be caused just by the general
distribution of galaxies in the sample. Hence, we applied the method described
above to discard this possible bias. The number of AGN was compared in the
region in common between the two samples defined by the intersection of regions
in which 95.45\% of the galaxies are within the area for each sample (equivalent
to 2 sigma in a normal distribution). This common region contains $\approx
206.8$ AMIGA galaxies and  $\approx 221.2$ HCG galaxies and is shown as a
contour on the panels of Fig.~\ref{fig:agn_comp}. Values will be measured within
this common region from now on. The fraction of Seyfert galaxies with respect to
the total number of AGN in HCG is 37\% \citep{Martinez2010}, while we find a
fraction of $\sim 30\%$ in AMIGA galaxies. Therefore, we do not expect to be
comparing very different types of AGN (or retired galaxies) between the two
samples.

Table~\ref{table:comp} shows the number of AGN in each sample. It also shows the
number of AGN that would be obtained by combining the galaxy distribution of
each sample with the AGN probability distribution of the other sample (as
defined in Sect.~\ref{sec:comp_method}). A binomial test comparing the observed
with the cross-estimated number of AGN galaxies yields p-values well above the
limit of $\alpha = 0.1$ for both samples. Hence, we cannot reject the hypothesis
that the AGN fractions are the same, after correcting for the luminosity and
morphology distribution. The lower panels of Fig.~\ref{fig:agn_comp} show the
cross-estimated distribution of AGN applying our method. The cross-estimated
distributions (e and f) are not very different from the observed ones (c and d
respectively), as is confirmed by the statistical tests. There is a slight
discrepancy in the distribution of AGN for early spirals between the observed
and estimated distribution in HCG (panels d and f) but the low number of
galaxies in this region makes this difference less significant. 

\begin{table}
\caption{Comparison between AMIGA and HCG samples.$^{\rm 1}$}
\label{table:comp}
~\\
\centering
\begin{tabular}{llccc}
\hline
 & & \multicolumn{2}{l}{Number of objects} & p val.$^{\rm 2}$ \\
 & & \multicolumn{1}{l}{p AMIGA} & \multicolumn{1}{l}{p HCG} & \\
\hline
\hline
AGN & AMIGA & $44.2 \pm 5.8$ & $\mathit{52.2 \pm 6.0}$ & $52.2\%$ \\
 & & (Fig.~\ref{fig:agn_comp}c)& (Fig.~\ref{fig:agn_comp}e) & \\
 & HCG   & $\mathit{56.5 \pm 6.2}$ & $56.0 \pm 6.4$ & $46.0\%$ \\
 & & (Fig.~\ref{fig:agn_comp}f)& (Fig.~\ref{fig:agn_comp}d) & \\
\hline
AGN + TO & AMIGA & $79.3 \pm 6.9$ & $\mathit{108.2 \pm 7.1}$ & $5.9\%$ \\
 & HCG & $\mathit{95.5 \pm 7.2}$ & $100.9 \pm 7.4$ & $16.7\%$ \\
\hline
SFN & AMIGA & $115.9 \pm 7.1$ & $\mathit{98.1 \pm 5.7}$ & $0.05\%$ \\
 & HCG & $\mathit{71.6 \pm 5.7}$ & $43.6 \pm 5.2$ & $6.9 \times 10^{-7}\%$ \\
\hline
\end{tabular}
\begin{list}{}{}
\item[$^{\rm 1}$] The rows correspond to the distribution of galaxies (sample)
and the columns to the probability of harbouring an AGN, an AGN + TO or a SFN.
The numbers in italics are the cross-estimated, calculated using the presented
method.
\item[$^{\rm 2}$] The p-value is computed using a binomial test for a given
probability and two different distributions.
\end{list}
\end{table}

We also compared the prevalence of AGN in both samples using a $\chi^{2}$ test.
We were limited by the relatively low number of AGN in both samples, therefore,
the bins were selected to allow a significant number ($n>6$) of observed and
expected AGN in each one: a) Morphology from E to Sa and $\log{(L_{\rm{B}})}$ from
9.5 to 11.0, b) morphology from Sab to Sd and $\log{(L_{\rm{B}})}$ from 9.5 to
11.0, and c) the region out of the other two bins. The bins are shown in the
left-middle panel of Fig.~\ref{fig:agn_comp}. The p-values for the comparisons
are 0.181 and 0.165, both above the 0.1 limit, hence, AGN distributions in AMIGA
and HCG are indistinguishable.

If we take into account both AGN and TOs, the p-value decreases (see
Table~\ref{table:comp}) and there is increasing evidence of a difference between
the distributions. This difference could be explained by the higher weight of
the star formation component in TOs.

Finally, the fraction of SFN nuclei in both samples was compared. There is
evidence of a low star formation rate
\citep{Verdes-Montenegro1998,Iglesias1999,Coziol2000} or even suppression of
star formation \citep{Sulentic2001,Durbala2008b} in compact groups. We checked
if the rate of star formation was different after correcting for the morphology
luminosity distribution. The result is shown in Table~\ref{table:comp}. In this
case, we can reject the hypothesis that the prevalence of SFN in both samples is
similar. We obtain always lower values in HCG even after the correction.

\section{Discussion and conclusions}
\label{sec:discon}

We presented a careful estimation of the optical nuclear activity in a
well-defined, statistically significant sample of isolated galaxies using SDSS
data. We aimed to determine the prevalence and properties of nuclear activity in
galaxies not affected by major tidal interactions during the last part of their
lifetime. To shed light on the effect of interaction on nuclear
activity, we compared our sample with a sample of galaxies that are strongly
affected by interaction (HCG galaxies).

From the 6th Data Release of the Sloan Digital Sky Survey, we obtained spectral
data that were inspected in a semi-automatic way to ensure the accuracy of the
data. We subtracted from the spectra the underlying stellar populations obtained
using the software Starlight and fitted the emission lines. We applied the
typical diagnostics diagrams to classify the type of nuclear emission using a
classification scheme that takes into account censored data. A catalogue of
spectroscopic data, stellar populations, emission lines and optical nuclear
activity classification was provided. The entire intermediate data are released
to allow the use of different nuclear activity classification criteria in the
future. Finally, we presented a method to compare the prevalence of AGN with
galaxies in denser environments that takes into account the effect of the
density-morphology/luminosity relations to avoid a bias in the comparison. The
prevalence of AGN and SFN was compared with galaxies in Hickson Compact Groups.

From the present study we found that
   \begin{enumerate}
      \item Our nuclear activity classification method, which takes into account
the information carried by censored data, seems to be robust with respect to
an increase of the noise level.
      \item The prevalence of optical nuclear activity in AMIGA isolated
galaxies is 20.4\%. This percentage considers Seyfert 2, Seyfert 1, LINERs, and
unclassified NLAGN (LINER or Seyfert 2). If we consider these types together
with TOs, the percentage
rises to 36.7\%.
      \item The fraction of AGN increases steeply towards higher $L_{\rm{K}}$ 
(proxy for stellar mass)
or $L_{\rm{B}}$ and earlier morphological types. 
      \item We found no evidence of a difference between
the population of AGN in AMIGA galaxies and in Hickson Compact Groups after
correcting for the effect of the
density-morphology/luminosity. On the other hand, we found significant evidence
of a difference for SFN with a higher prevalence in isolated galaxies with
respect to galaxies in HCGs, which is consistent with a lowered SF in HCG.
   \end{enumerate}

The isolation criteria for AMIGA galaxies imply nearest-neighbour crossing times
of at least 3 Ga \citep[in prep.]{Verdes-Montenegro2005,Verley2007b,Mamen2012}
That fact, and the general properties of the sample, lead us to consider that
AMIGA galaxies were not affected by major tidal interactions during the last
part of their lifetime. The time-scale of the duty cycle of an AGN is very short
\citep[less than 100 Ma;][]{Haehnelt1998,Greene2007,Ho2008,Schawinski2009} in
comparison with the isolation time of our galaxies. Hence, AGN activity in AMIGA
galaxies should be unrelated to a major interaction event.

Major interactions have been proposed as a mechanism to trigger AGN activity in
some cases \citep[see review by][]{Combes2003,Ellison2011,Liu2012}, although no
clear relation was found in some studies using the SDSS \citep{Li2006,Li2008}.
Seyferts do not \textit{require} external interactions to be fueled
\citep[e.g.][]{Combes2003}, thus, alternative mechanisms such as minor mergers,
accretion of gas from cosmic filaments and/or internal mechanisms
\citep{Combes2006} are needed to explain the presence of optical active nuclei
in isolated galaxies. On the other hand, these alternative mechanisms do not
trigger radio nuclear activity in isolated galaxies and an additional factor,
linked to the environment (directly or indirectly), is needed to explain the
presence of a radio jet \citep{Sabater2008,TesisMia,Sabater2010p}. 

If we consider a traditional feeding mechanism of black holes driven by major
mergers \citep{Kormendy2011}, the relatively high fraction of AGN in isolated
galaxies, dominated by pseudo-bulges \citep{Durbala2009}, might favour an
alternative local feeding mechanism driven by the slow accretion of cold gas  
\citep{Hopkins2006}. This mechanism is probably closer related to the activity
found in galaxies with pseudo-bulges \citep{Jiang2011}. This could mean that the
black holes found in AMIGA galaxies are close to the black hole seeds proposed
by \citet{Kormendy2011}, the seeds of the larger black holes that are grown by
mergers.

The comparison with galaxies in compact groups yielded that the difference in
the fraction of AGN between isolated galaxies and those in compact groups is not
significant (after correcting for the morphology and luminosity distribution),
while for SFN it is clearly significant. However, we recall that we are
comparing fractions and not intensities or types of SF or AGN activity. One
possible explanation is that the expected increase of AGN activity in HCG due to
interactions \citep{Ellison2011,Liu2012} could be compensated for by a decrease
due to the environmental density
\citep{Kauffmann2004,vonderLinden2010,Gavazzi2011}. An additional consideration
that has to be taken into account is the possible contamination of the different
samples by retired galaxies, whose nuclear emission resembles the emission of
weak AGN (usually LINERs) but is actually powered by low-mass evolved stars
\citep{CidFernandes2010,Cid-Fernandes2011}. To clarify those points, the results
of this paper will be extended in the future to study the independent effects of
environment and interaction on the    triggering of both optical and radio
nuclear activity in a large homogeneously selected sample of SDSS galaxies.

Even though isolated galaxies do not present radio nuclear activity, the
prevalence of optical nuclear activity is not negligible. Indeed, this leads us
to conclude that a major interaction is not a necessary condition for the
triggering of optical nuclear activity. Hence, at least part of the growth of a
black hole could be produced without the need of a major merging event.

\begin{acknowledgements}
We would like to thank the anonymous referee for numerous suggestions that
improved the clarity of the paper.
JSM, LVM, SL and JS were partially supported
by DGI Grants 
AYA 2005-07516-C02-01, AYA 2008-06181-C02-01 and Junta de Andaluc\'{\i}a
(Spain) Grants TIC-114 and F08-FQM-4205.
Funding for the SDSS and SDSS-II has been provided by the Alfred P. Sloan
Foundation, the Participating Institutions, the National Science Foundation, the
U.S. Department of Energy, the National Aeronautics and Space Administration,
the Japanese Monbukagakusho, the Max Planck Society, and the Higher Education
Funding Council for England. The SDSS Web Site is http://www.sdss.org/.
The SDSS is managed by the Astrophysical Research Consortium for the
Participating Institutions. The Participating Institutions are the American
Museum of Natural History, Astrophysical Institute Potsdam, University of Basel,
University of Cambridge, Case Western Reserve University, University of Chicago,
Drexel University, Fermilab, the Institute for Advanced Study, the Japan
Participation Group, Johns Hopkins University, the Joint Institute for Nuclear
Astrophysics, the Kavli Institute for Particle Astrophysics and Cosmology, the
Korean Scientist Group, the Chinese Academy of Sciences (LAMOST), Los Alamos
National Laboratory, the Max-Planck-Institute for Astronomy (MPIA), the
Max-Planck-Institute for Astrophysics (MPA), New Mexico State University, Ohio
State University, University of Pittsburgh, University of Portsmouth, Princeton
University, the United States Naval Observatory, and the University of
Washington. 
This publication makes use of data products from the Two Micron All Sky Survey,
which is a joint project of the University of Massachusetts and the Infrared
Processing and Analysis Center/California Institute of Technology, funded by the
National Aeronautics and Space Administration and the National Science
Foundation.
We acknowledge the usage of the HyperLeda database (http://leda.univ-lyon1.fr)
\end{acknowledgements}

\bibliographystyle{aa}
\bibliography{databasesol}

\Online
\begin{appendix}
\section{Additional data}
There are some additional data associated with this study: a) SDSS photometric
data that were processed but not used for this study and b) Hyperleda data that
were compiled for the comparison presented in Sect.~\ref{sec:comp}. These data
are presented here and in http://amiga.iaa.es/ for future reference and to allow
the data replication and reproducibility.

\begin{figure}
   \centering
   \includegraphics[width=0.48\textwidth]{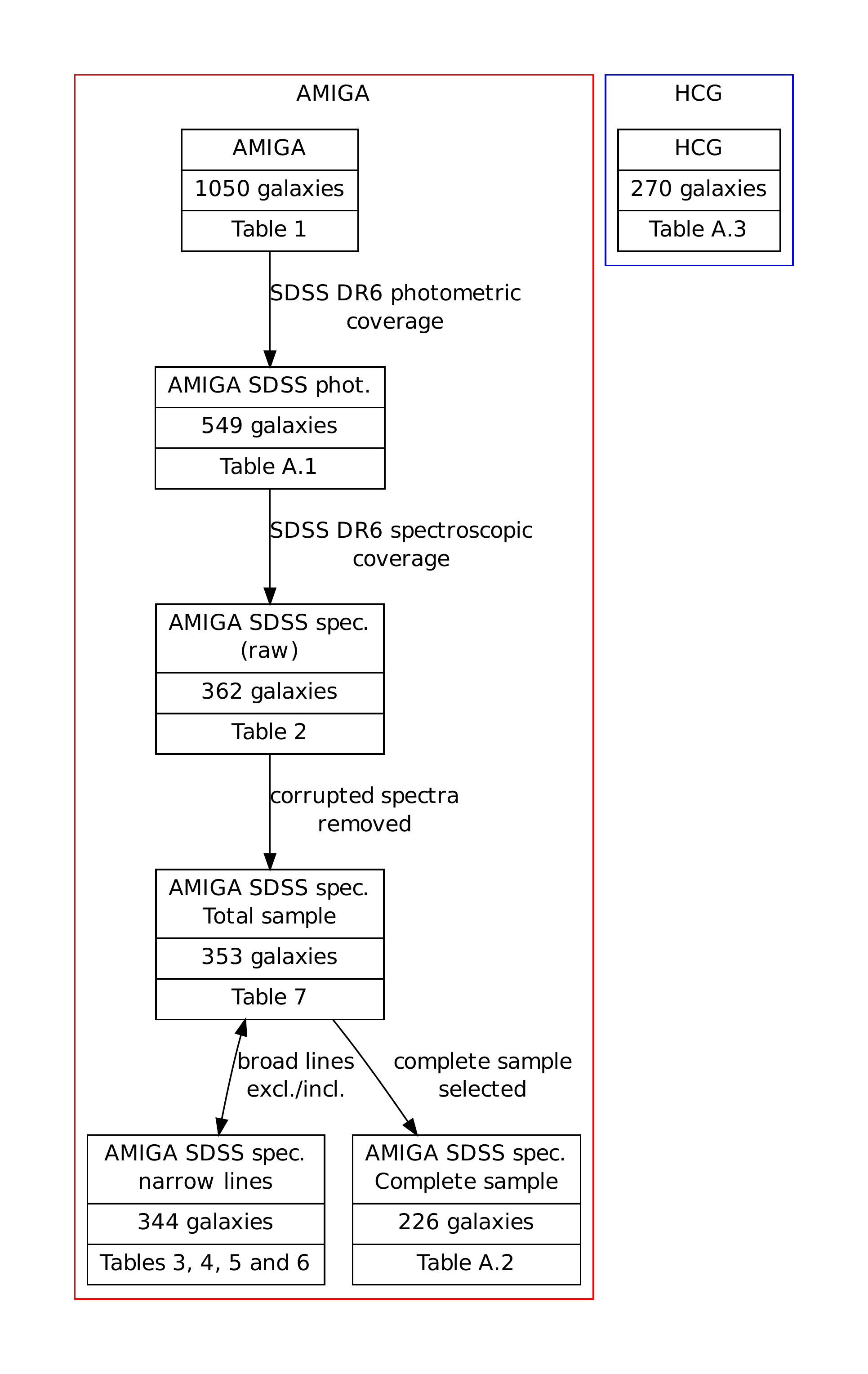}
   \caption{Description of the samples and data presented in the
different
tables within the paper.}
   \label{fig:samples}
\end{figure}

A diagram of the samples and data presented in the paper is shown in
Fig.~\ref{fig:samples}.

\subsection{SDSS photometric data}

We cross-correlated the positions of our galaxies \citep{Leon2003} with the SDSS
DR6 photometric catalogue using the interface available on the SDSS web-page
(search radius of 0.5\arcmin). We obtained a list of photometric sources near
the estimated centre of each galaxy as well as the corresponding SDSS spectral
data when available. The right objects were identified using a semi-automatic
selection algorithm selecting the object with the highest \textit{z} band flux
followed by a visual inspection as a cross-check. Only $\sim 8\%$ of the
photometric objects were clearly misclassified with the usual source of
confusion involving a nearby star. A total of 549 AMIGA galaxies were included
in DR6.

In six cases the photometry for the galaxy was assigned to an object whose
estimated position was shifted from the centre of the galaxy. In these cases, we
manually chose the correct photometric object. We inspected in more detail some
special cases that were finally discarded from our catalogue (CIG~402 is
strongly affected by the light of a nearby star and CIG~802 and ~388 are
resolved Milky Way satellites). As a final check we compared the apparent
magnitudes in g-band from the SDSS data and in B-band from the AMIGA database
\citep{Verdes-Montenegro2005}. Although the filter shapes are different and the
shape of the spectral energy distribution of the galaxy directly affects their
ratio, a high difference between these two magnitudes may indicate an error in
the selection of the photometric data. The mean of the ratio is 1.044 and the
standard deviation is 0.070. In seven cases the difference was found to be above
$3\sigma$, suggesting that the SDSS photometric data might refer to a restricted
region rather than the entire galaxy. The photometric data for these cases were
flagged. The values for the SDSS photometry are presented in
Table~\ref{table:phot}.

\begin{table*}
\caption{Catalogue of SDSS photometry.$^1$}
\label{table:phot}
~\\
\centering
\begin{tabular}{ c c c c c c c r}
\hline
CIG & ObjId & u & g & r & i & z & comment \\
\hline \hline
11 & 588015508197212201 & 15.76 & 14.45 & 13.83 & 13.52 & 13.33 &  \\
12 & 588290879491408031 & 17.04 & 15.57 & 14.88 & 14.55 & 14.30 &  \\
16 & 587731187816857647 & 17.04 & 15.26 & 14.51 & 14.12 & 13.83 &  \\
19 & 587730773354479638 & 16.83 & 14.95 & 14.10 & 13.68 & 13.32 &  \\
33 & 587731186208669747 & 14.90 & 13.57 & 12.95 & 12.65 & 12.40 &  \\
 ... & ... & ... & ... & ... & ... & ... & ... \\
\hline
\end{tabular}
\begin{list}{}{}
\item[$^{1}$] AMIGA galaxies with SDSS DR6 photometry ($n=549$).
Columns:
(1) CIG catalogue number;
(2) SDSS ObjId of the selected SDSS photometric object. In some
cases it does not match the ObjId associated to the spectroscopic object.
(3) to (7) SDSS photometric magnitudes;
(8) comments for galaxies with poor photometry (see text).
\end{list}
\end{table*}

\subsection{Hyperleda data}

Hyperleda data were compiled for the AMIGA and HCG samples 
(see Sect.~\ref{sec:comp}). In Table~\ref{table:hl_amiga}, the
compiled morphological classification and absolute magnitude
for AMIGA galaxies are shown. In Table~\ref{table:hl_hcg},
the compiled morphological classification and absolute magnitude
plus the nuclear activity classification obtained from 
\citet{Martinez2010} are shown.

\begin{table}
\caption{Hyperleda data for AMIGA galaxies.$^1$}
\label{table:hl_amiga}
~\\
\centering
\begin{tabular}{ c c c }
\hline
CIG & Morpho. & $M_{\mathrm{B}}$ \\
\hline \hline
 11 & 5.1 & $-20.132$\\
 56 & 3.4 & $-20.641$ \\
60 & 2.8 & $-19.503$ \\
187 & 3.9 & $-20.805$ \\
189 & $-5.0$ & $-19.511$ \\
 ... & ... & ...\\
\hline
\end{tabular}
\begin{list}{}{}
\item[$^{1}$] \textit{SDSS complete sample} ($n=226$). Columns:
(1) CIG catalogue number;
(2) morphological classification (RC3);
(3) absolute magnitude in B-band.
\end{list}
\end{table}

\begin{table}
\caption{Hyperleda data for HCG galaxies.$^1$}
\label{table:hl_hcg}
~\\
\centering
\begin{tabular}{ l c c c }
\hline
HCG & Morpho. & $M_{\mathrm{B}}$ & class.\\
\hline \hline
H01a & 4.6 & $-21.768$ & SFN \\
H01b & $-2.9$ & $-20.564$ & - \\
H01c & $-2.3$ & $-20.214$ & AGN \\
H01d & $-1.5$ & $-19.155$ & AGN \\
H03a & 4.3 & $-20.635$ & AGN \\
 ... & ... & ... & ...\\
\hline
\end{tabular}
\begin{list}{}{}
\item[$^{1}$] HCG galaxies ($n=270$). Columns:
(1) HCG catalogue name;
(2) morphological classification (RC3);
(3) absolute magnitude in B-band;
(4) nuclear activity classification. 
  Possible values: AGN, SFN, TO, AGN/TO or unclassified (-).
\end{list}
\end{table}

\end{appendix}

\end{document}